\documentclass[]{pasj02} 
\usepackage[switch,mathlines]{lineno} 
\usepackage{booktabs}
\usepackage{url}

\jyear{2024}
\Received{}
\Accepted{}

\begin{document} 

\title{Spectroscopic diagnostics of high-temperature plasma in stellar corona using Fe \emissiontype{XXIV}--\emissiontype{XXVI} K-shell lines with XRISM}

\author{
Miki~\textsc{Kurihara},\altaffilmark{1,2}\altemailmark\orcid{0000-0002-3133-9053} \email{kurihara@ac.jaxa.jp, miki39kurihara@g.ecc.u-tokyo.ac.jp} 
Masahiro~\textsc{Tsujimoto},\altaffilmark{2}\orcid{0000-0002-9184-5556}
Marc~\textsc{Audard},\altaffilmark{3}\orcid{0000-0003-4721-034X}
Ehud~\textsc{Behar},\altaffilmark{4}\orcid{0000-0001-9735-4873}
Liyi~\textsc{Gu},\altaffilmark{5}\orcid{0000-0001-9911-7038}
Kenji~\textsc{Hamaguchi},\altaffilmark{6,7}\orcid{0000-0001-7515-2779}
Natalie~\textsc{Hell},\altaffilmark{8} \orcid{0000-0003-3057-1536}
Caroline~A.~\textsc{Kilbourne},\altaffilmark{6}\orcid{0000-0001-9464-4103}
Yoshitomo~\textsc{Maeda},\altaffilmark{2}\orcid{0000-0002-9099-5755}
Frederick~S.~\textsc{Porter},\altaffilmark{6}\orcid{0000-0002-6374-1119}
Haruka~\textsc{Sugai},\altaffilmark{9} 
\and
Yohko~\textsc{Tsuboi}\altaffilmark{9}\orcid{0000-0001-9943-0024}
}
\altaffiltext{1}{Department of Astronomy, Graduate School of Science, The University of Tokyo, Bunkyo-ku, Tokyo 113-0033, Japan}
\altaffiltext{2}{Japan Aerospace Exploration Agency, Institute of Space and Astronautical Science, Chuo-ku, Sagamihara, Kanagawa 252-5210, Japan}
\altaffiltext{3}{Department of Astronomy, University of Geneva, CH-1290 Versoix, Switzerland}
\altaffiltext{4}{Department of Physics, Technion, Technion City, Haifa 3200003, Israel}
\altaffiltext{5}{SRON Netherlands Institute for Space Research, Niels Bohrweg 4, 2333 CA Leiden, The Netherlands}
\altaffiltext{6}{NASA's Goddard Space Flight Center, Greenbelt, MD 20771, USA}
\altaffiltext{7}{Department of Physics, University of Maryland, Baltimore County, Baltimore, MD 21250, USA}
\altaffiltext{8}{Lawrence Livermore National Laboratory, CA 94550, USA}
\altaffiltext{9}{Department of Physics, Faculty of Science and Engineering, Chuo University, Bunkyo-ku, Tokyo 112-8551, Japan}

\KeyWords{atomic processes, stars: coronae, stars: individual (GT Mus), techniques: spectroscopic, X-rays: stars}

\maketitle

\begin{abstract}
 The RS CVn type binary star GT Mus was observed during its quiescence using the
 \textit{Resolve} X-ray microcalorimeter spectrometer onboard XRISM. The main and
 satellite lines of the Fe \emissiontype{XXIV}--\emissiontype{XXVI} K-shell transitions
 were resolved for the first time from stellar sources. We conducted line ratio analysis
 to investigate any deviations from collisional ionization equilibrium (CIE) and Maxwell
 electron energy distribution with a single-temperature. By using five combinations of
 direct excitation lines and dielectronic recombination satellite lines in three line
 complexes (Fe He$\alpha$, Ly$\alpha$, and He$\beta$), we found that the plasma is well
 characterized by two-temperature thermal plasmas with temperatures of 1.7 and 4.3 keV,
 which is consistent with a thermal broadening of Fe \emissiontype{XXV} and the
 broadband fitting results in the 1.7--10 keV band. Other forms of deviation from a
 single-temperature plasma, such as different ionization and electron temperatures or
 the $\kappa$ distribution for the electron energy distributions, are not favored, which
 is reasonable for stellar coronae at quiescence. This study demonstrates the utility of
 the Fe K-shell line ratio diagnostics to probe plasma conditions using X-ray
 microcalorimeters.
\end{abstract}


\section{Introduction}\label{s1}
Collisional plasmas in the Universe often deviate from the thermal equilibrium of a
single-temperature. The deviation provides insights into the spatial distribution, the
dynamics, and the mechanism of heating of such plasmas through the equilibration
processes. The deviation can take several different forms, including the
multi-temperature distribution, the suprathermal electron energy distribution, and
non-equilibrium ionization (NEI). The multi-temperature distributions are often seen in
systems of different plasma sources integrated in the line of sight. Suprathermal
electrons represented by an additional higher-energy population upon the thermal
Maxwellian distribution are also seen in in-situ observations of heliospheric plasmas
(e.g., \cite{maksimovic1997}). NEI plasmas occur when ions are either under-ionized or
over-ionized relative to the electron temperature, which is a feature ubiquitously seen
in tenuous plasmas of an impulsive energy injection such as supernova remnants (e.g.,
\cite{vink2020,yamaguchi2022}).

X-ray line ratio measurements are a promising method for probing these departures from a
thermal equilibrium. Deviated conditions influence the ionization, recombination, and
excitation rates, which in turn affect the charge and level populations, thereby
modifying the intensity of emission lines. For multi-temperature and suprathermal plasma
diagnostics, \citet{gabriel1972,gabriel1979} first proposed using lines formed by the
dielectronic recombination (DR) process, which are sensitive to variations in electron
energy distributions. For NEI diagnostics, the ratio of strong direct excitation (DE)
lines from ions of different charge states is useful.

The Fe K-shell line complex in the 6--9~keV provides particularly rich information for
several reasons. Fe has the largest atomic number ($Z$) among abundant elements in the
Universe. The K-shell line energies increase with $Z$ and those of Fe probe the plasma
of temperatures in 1--10~keV (see \cite{hell2020} for a review). The DR to DE line ratio
scales roughly as $Z^4$ \citep{gabriel1972_2}, thus the weaker DR lines are more
accessible as $Z$ increases. The ionization parameter $\tau = \int dt~n_{e}$, where
$n_e$ is electron density, is a measure of the Coulomb relaxation time scale of NEI
plasmas.  Its $Z$-dependency was investigated for major elements from C to Ni
\citep{smith2010}. In the 1--10~keV temperature range, the time to reach an equilibrium
is longer for higher $Z$ elements than lower $Z$ elements, which were studied using the
X-ray grating spectrometers onboard Chandra and XMM-Newton (e.g.,
\cite{nordon2007,audard2003}).

In order to take advantage of the Fe K-shell lines, a spectral resolving power $R \equiv
E / \Delta E \gtrsim 1000$ is required for the energy band $E \gtrsim 6.4$ keV. Until
the advent of the X-ray microcalorimeter spectrometers, this was achieved only in some
selected narrow energy bands for solar X-ray observations using rotating or bent crystal
spectrometers onboard P78-1, Hinotori (Astro-A), Yohkoh (Solar-A), and SMM
\citep{doschek1979,doschek1980,feldman1980,dubau1981,parmar1981,tanaka1982,tanaka1986,watanabe2024},
The first X-ray microcalorimeter observation was made using the Soft X-ray Spectrometer
(SXS; \cite{kelley2016}) onboard Hitomi (Astro-H; \cite{takahashi2018}) for the Perseus
cluster of galaxies \citep{hitomi2018}. Despite its short lifetime, the SXS demonstrated
the capability of X-ray microcalorimeters for diagnosing celestial plasmas using Fe
K-shell lines over a wide energy range covering higher series ($n \rightarrow 1$; $n \ge
2$) lines \citep{aharonian2018}.

The unique capability of the SXS is now recovered with the \textit{Resolve} instrument
\citep{ishisaki2022} onboard the X-ray Imaging and Spectroscopy Mission (XRISM;
\cite{tashiro2020}). The instrument has been yielding high-resolution Fe K-shell spectra
of collisionally-ionized plasmas from interstellar gas \citep{xrism2025o}, supernova
remnants \citep{xrism2024}, and cluster of galaxies \citep{xrism2025,audard2025}. Stars
are an important category for applying plasma diagnostics for their relatively slow bulk and
turbulent motions, which make interpretation easier, as well as their rapid dynamical
developments on human time scales during flares. Comparison to the Sun (e.g.,
\cite{huenemoerder2013}) is also unique to stellar sources.

In this paper, we present the first results of the Fe K-shell line diagnostics of
stellar coronae using an X-ray microcalorimeter. Line ratio analysis is performed to
investigate any deviations from single-temperature plasmas. In addition to modeling the
Fe He$\alpha$ and Ly$\alpha$ complexes as in solar studies, we also use the Fe He$\beta$
complex at 7.781 keV, which was observed with $R > 1000$ for the first time in stars
including the Sun \citep{phillips2012}.

\medskip

The outline of this paper is as follows. In \S~\ref{s2}, we describe the target, XRISM
observation, and data reduction. In \S~\ref{s3}, we present the \textit{Resolve} light
curve and X-ray spectrum, followed by the results of the spectral fitting analysis. In
\S~\ref{s4}, we apply the line ratio techniques and discuss the plasma properties. In
\S~\ref{s5}, we summarize the findings and provide the conclusion. Throughout the
paper, the quoted uncertainties are for 90\% statistical error.

\section{Observations}\label{s2}
\begin{figure*}[!hbtp]
 \begin{center}
 \includegraphics[width=0.99\linewidth]{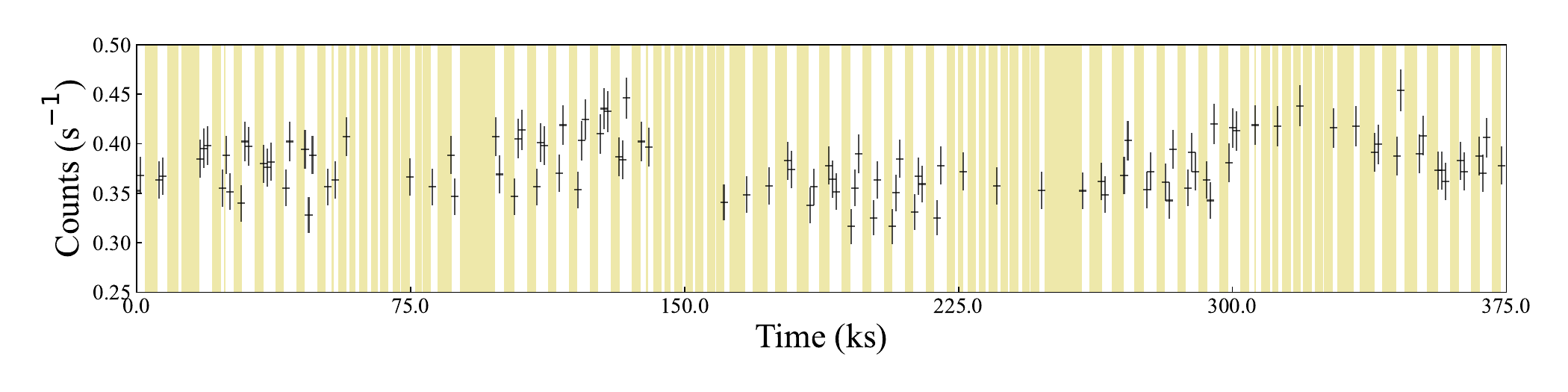}
 \end{center}
 \caption{\textit{Resolve} light curve in the 1.7--10.0 keV with a 1024~s binning
 (bottom) after the standard screening. The time origin is 2024-08-13T02:12:12. The
 observation intervals are shown with a white background.
 Alt text: X-ray light curve in the 1.7–-10.0 keV energy range. 
 }
 \label{f01}
\end{figure*}

\subsection{Target}\label{s2-1}
GT Mus is the target of this study. It is a quadruple system at a distance of
$\sim$141~pc by a parallax measurement with Gaia \citep{prusti2016,vallenari2023}. It
consists of two binaries: HD101379 and HD101380. HD101379 is a single-line spectroscopic
binary composed of a G5 and a G8 giant, while HD101380 is an eclipsing double-line
spectroscopic binary composed of an A0 and an A2 dwarf. The orbital periods of HD101379
and HD101380 are 61.4 and 2.7546 days, respectively \citep{murdoch1995,collier1982}. The
former is an RS-CVn type source and is considered to be the primary source of X-ray
emission.

GT Mus is known to exhibit giant flares, some of which were captured in X-ray
observations. The gas proportional counter (GSC; \cite{mihara2011}) onboard the Monitor
of All-sky X-ray Image (MAXI; \cite{matsuoka2009}) detected eleven flares in eight years
with a peak flux of 34--130 mCrab and the released energy in a single flare amounts to
$\sim 10^7$ times larger than that of X-class flares of the Sun
\citep{tsuboi2016,sasaki2021}. The highest energy photons were detected from GT Mus
among all stellar flares using the INTEGRAL \citep{winkler2003} soft gamma-ray imager (ISGRI;
\cite{lebrun2003}) up to 40 keV during a flare \citep{sguera2016}. Such giant flares are
rare, and their peaks are difficult to capture by X-ray telescopes with a small field of
view.

The X-ray Imaging Spectrometer (XIS; \cite{koyama2007}) onboard Suzaku (Astro-E2;
\cite{mitsuda2007}) and the Neutron star Interior Composition ExploreR (NICER;
\cite{gendreau2016}) observed the decaying phase of flares
\citep{eze2022,sasaki2021,xu2016} and recorded thermal plasma emission with a
temperature of $\sim$5~keV. Some flares last sufficiently long enough up to a few days
to fall within the XRISM target-of-opportunity response time. These flares are expected
to yield count rates in the appropriate dynamic range of the \textit{Resolve} X-ray
microcalorimeter. Although no such observations have been made as of writing, we
anticipate a few during the mission.

Even during the quiescent phase, GT Mus is an attractive target. While other stars
exhibit soft X-ray emission of a $\lesssim 2$~keV temperature from their stellar coronae
(e.g., \cite{sanz2002,audard2003a,pandey2012}), GT Mus exhibits an exceptionally hard
X-ray emission with a $\sim$5~keV temperature even without giant flares
\citep{xu2016,sasaki2021}.  It is not well understood for the cause of its exceptional
behavior, but this provides a unique opportunity for the Fe K band spectroscopy of
stellar coronae using an X-ray microcalorimeter unconstrained by unpredictable flare
behaviors. This is what we present in this paper.

\subsection{Instrument}\label{s2-2}
XRISM hosts two scientific instruments that operate simultaneously. One is the X-ray
microcalorimeter spectrometer $\textit{Resolve}$ \citep{ishisaki2022} and the other is
the X-ray CCD imaging spectrometer $\textit{Xtend}$ \citep{mori2022,noda2025}. In this paper, we
focus on the former.

$\textit{Resolve}$ is an X-ray spectrometer based on X-ray microcalorimetry
\citep{mccammon1984}. The detector array consists of 6$\times$6 pixels, with each pixel
containing a HgTe X-ray absorber and an ion-doped Si thermister thermally anchored to the
50~mK stage \citep{kilbourne2018d} controlled by the adiabatic demagnetization refrigerator
(ADR; \cite{shirron2018}). One of the 36 pixels is displaced from the array for
calibration purposes. 
The in-orbit performance of energy resolution and absolute energy scale has been evaluated to be $\Delta E \sim
4.5$~eV (FWHM) and $\delta E \sim 0.3$~eV in the Fe K band \citep{porter2024,eckart2024}. The
exception is the bandpass; the lower energy range is currently limited to above 1.7~keV
due to the cryostat transmissive window \citep{midooka2021}, which is yet to be opened.

For the purpose of this study, several unique characteristics of \textit{Resolve} provide
distinct advantages. One is the energy resolution $R = 1300$ at 6 keV, better at higher energies, which enables the
separation of fine-structure levels and satellite lines of Fe \emissiontype{XXV}
He$\alpha$ ($\sim$6.7~keV) and Fe \emissiontype{XXVI} Ly$\alpha$ (7.0~keV) line
complexes. Another advantage is the high-energy end of the bandpass, which is achieved
by the HgTe absorbers instead of the Si absorbers in conventional X-ray
spectrometers. Together with a high throughput and low background rate, this allows the
detection of higher-series lines such as Fe \emissiontype{XXV} He$\beta$ at
7.8~keV. X-ray spectrometers with such unique features have not been achieved even in
solar observations.

\subsection{Observation and data reduction}\label{s2-3}
GT Mus was observed as one of the performance verification phase targets (sequence
number 300000010) from 2024 August 13 02:03:40 to August 17 10:12:28 (UT). The total
telescope time of 375~ks was interrupted by (i) occultation of the target by the Earth,
(ii) XRISM passages through the South Atlantic Anomaly (SAA) region, (iii) ADR recycling, and
(iv) $^{55}$Fe source illumination for calibration.
As a result, the observing efficiency was 53\%, yielding an effective exposure time of
199~ks (figure\ref{f01}).

The \textit{Resolve} energy scale was calibrated both on the ground and using on-board calibration sources in-flight \citep{eckart2024}. The time dependent detector gain directly affects the energy scale and was corrected during the observation of GT Mus using periodic observations of $^{55}$Fe X-ray sources installed on the instrument filter wheel (FW) assembly. The sources were rotated into the field of view (FOV) during several earth occultations of the celestial target. For this observation, there were twelve, $\sim$30 minute, observations of the FW sources that were then used to correct the time dependent detector gain in the standard process for \textit{Resolve} \citep{porter2016,porter2024}. \textit{Resolve} also contains a calibration pixel that is located on the main detector array but just outside of the FOV. The calibration pixel is used to monitor the efficacy of the gain correction during each observation, outside of the fiducial measurements, since it is illuminated continuously by a dedicated $^{55}$Fe X-ray source. For this observation, the energy scale reconstruction was very precise, with the calibration pixel yielding an accuracy of $-0.228 \pm 0.006$ eV at 5.9 keV. This is added in quadrature to the systematic energy scale uncertainties of 0.3 eV from the in-flight calibration to yield an energy scale uncertainty in the region from 5.4--8 keV that is covered by the on-board calibration sources.  This yields an estimated uncertainty for this observation of 0.38 eV, 5.4--8 keV. The estimated systematic energy scale uncertainty is $\sim$1 eV from the bottom of the band to 5.4 keV, and $\sim$2 eV above 8 keV. We used the standard \textit{Resolve} redistribution response matrix file (RMF) generator for this observation that includes the energy dependent core line spread function (energy resolution) that varies by pixel. The line spread function for \textit{Resolve} has been verified in flight, and also using the FW fiducial calibrations during this observation. The standard systematic uncertainty in the core line spread function for \textit{Resolve} is 0.13 eV at 6 keV and 0.17 eV at 7 keV.


We started with the products of the standard pipeline version 03.00.011.008
\citep{doyle2022}. We combined all 35 pixels in the array and used high-primary (Hp)
grade events only, which are temporally isolated from other events in the same pixel and
are well calibrated for high-resolution spectroscopy. We applied further
screening to reduce background events based on X-ray pulse rise time and energy
\citep{mochizuki2025}. The fraction of the Hp grade events is 0.96, and the resultant
mean count rate is 0.38~s$^{-1}$ in the 1.7--12~keV band. This count rate is low enough
to disregard artifacts caused by high count rates \citep{mizumoto2025}. The background
rate is also negligible in the energy range of interest \citep{kilbourne2018}.

We used \texttt{HEASoft} version 6.34.0 for data analysis. The spectrum was
optimally binned \citep{kaastra2016}, with a minimum of 10 counts per bin. The telescope
and detector response files were generated using the \texttt{xaarfgen} and
\texttt{rslmkrmf} tasks, respectively. For the detector response, we included
off-diagonal components in the redistribution matrix to account for escape peaks, Si
fluorescence, and electron loss continuum, in addition to the diagonal line spread
function at the core. For the background, we used a model spectrum of the non-X-ray
background \footnote{\url{https://heasarc.gsfc.nasa.gov/docs/xrism/analysis/nxb/nxb_spectral_models.html}} and adjusted its normalization with a diagonal response to match the observed spectrum in the
14--17~keV range, where celestial counts are negligible.

\section{Analysis}\label{s3}
We present the X-ray light curve and spectrum in \S~\ref{s3-1}, and describe the entire
spectrum with a single-temperature plasma model (\S~\ref{s3-2-1}) and individual line
complexes with phenomenological models (\S~\ref{s3-2-2}).

\begin{figure*}[!hbtp]
 \begin{center}
 \includegraphics[width=0.99\linewidth]{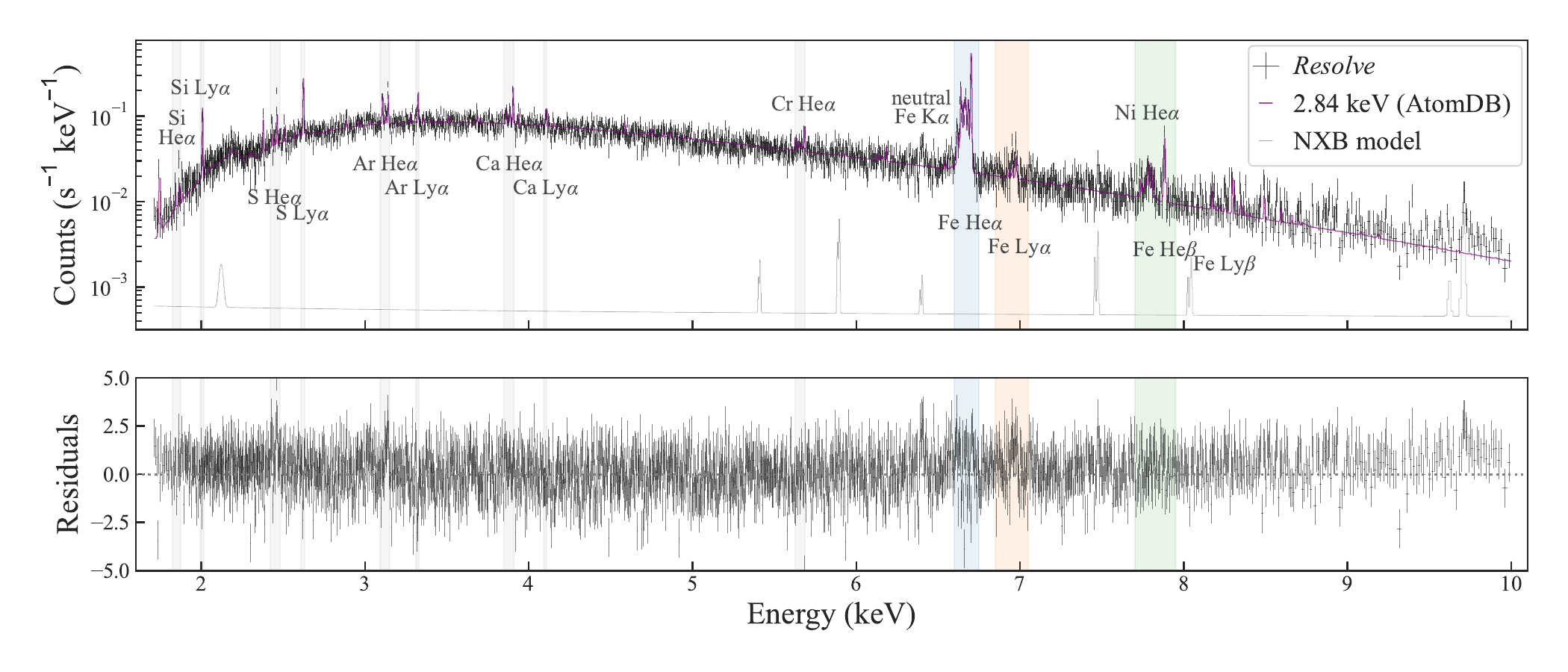}
 \end{center}
 \caption{(top) \textit{Resolve} spectrum in the 1.7--10.0 keV with optimal binning
 (black crosses) compared to the fiducial 2.84~keV thermal plasma model (purple curve)
 and the background model spectrum (grey). The blue, orange, and green stripes indicate the Fe
 He$\alpha$, Ly$\alpha$, and He$\beta$ line complexes, which we study in more detail in
 figure~\ref{f06}. (bottom) The residuals to the fiducial model fit.  Alt text: A
 two-panel figure showing an X-ray spectrum in the 1.7-–10.0 keV range.}  \label{f02}
\end{figure*}

\subsection{Light curve and spectrum}\label{s3-1}
Figure \ref{f01} shows the 1.7--10~keV light curve. The X-ray flux is consistent with
previous observation of GT Mus during quiescence \citep{sasaki2021}.  Given the small
variability and the observation duration being much shorter than the orbital period of the RS
CVn binary ($\sim 61.4$ day), we combined all events into a single spectrum without time
slicing.

Figure \ref{f02} shows the 1.7--10 keV spectrum. Numerous lines are recognized upon the
continuum, including the He$\alpha$ and Ly$\alpha$ lines of Si, S, Ar, Ca, Cr, Mn, Fe, and
Ni, as well as He$\beta$ and Ly$\beta$ lines of Fe. The K$\alpha$ fluorescence lines of
quasi-neutral Fe are also found. Notably, Fe K features are particularly prominent, with
the line complexes of Fe\emissiontype{XXV} He$\alpha$, Fe\emissiontype{XXVI} Ly$\alpha$,
and Fe\emissiontype{XXV} He$\beta$ clearly resolved. A comparison with the background
spectrum confirms that its contamination is negligible except for Au L lines at 9.7~keV.

\subsection{Spectral modeling}\label{s3-2}
\subsubsection{Broadband}\label{s3-2-1}
We first constructed the fiducial model for the broadband spectrum using the 1.7--10 keV
band source spectrum. The background spectrum was not subtracted, as its contribution is
negligible with $\lesssim 10$\%. Spectral fitting was performed using the \texttt{xspec}
software based on $\chi^2$ statistics \citep{arnaud1996}.

The model consists of a single-temperature thermal plasma emission (\texttt{apec} model;
\cite{smith2001}) attenuated by the X-ray photoelectric absorption by the interstellar
matter \citep{wilms2000}. The chemical abundance table of \citet{anders1989} was
used. The best-fit values of free parameters are the plasma temperature ($kT =
2.84^{+0.04}_{-0.03}$ keV) and the volume emission measure ($EM =
31.4^{+0.5}_{-0.5}\times 10^{54}$~cm$^{-3}$), the relative abundance of Fe
($Z_{\mathrm{Fe}} = 0.25 ^{+0.01}_{-0.01}~Z_{\odot}$), Ni ($Z_{\mathrm{Ni}}= 0.41
^{+0.16}_{-0.15}~Z_{\odot}$), and the other major elements ($0.36-0.7~Z_{\odot}$), and
the interstellar absorption column density ($N_{\mathrm{H}} =
1.0\times10^{14}~^{+8.2\times10^{21}}_{-1.0\times10^{14}}$~cm$^{-2}$). A satisfactory result was obtained with a reduced $\chi^2
= 2429.56 / 2169$, which describes the overall spectrum well (figure~\ref{f02}),
although some residuals remain in the Fe K complexes. In the broadband fitting using the
$\chi^{2}$ statistics, informative line features are not fully utilized, as spectral
bins with continuum emission dominate the statistics. We further examine these line
complexes using a different approach below.

\subsubsection{Line complexes}\label{s3-2-2}
We describe the line complexes using a phenomenological model consisting of individual
emission lines and a continuum. Three energy bands, colored in figure \ref{f02}, are
fitted separately: Fe\emissiontype{XXV} He$\alpha$ (blue; 6.603--6.750 keV),
Fe\emissiontype{XXVI} Ly$\alpha$ (orange; 6.850--7.050 keV), and Fe\emissiontype{XXV}
He$\beta$ (green; 7.700--7.950 keV). The lines used for the phenomenological fittings
are summarized in table \ref{t01}.

\paragraph{Models}
The model for the Fe He$\alpha$ complex was constructed as follows. It is composed of
main lines of Fe \emissiontype{XXV} and satellite lines of Fe
\emissiontype{XXIII}--\emissiontype{XXIV}. We selected 16 strongest lines based on the
\texttt{Chianti} code \citep{dere1997,dere2023} for the 2.84 keV thermal plasma
(\S~\ref{s3-2-1}). We confirmed that the dependence on the electron density
($n_{\mathrm{e}}$) is negligible for $n_\mathrm{e} \leq 10^{14}$~cm$^{-3}$. The number
of selected lines was determined considering the blending by non-selected weaker lines
and the data statistics, referring to the previous ground measurements
\citep{decaux1997,beiersdorfer1993,gu2012} and solar observation (e.g.,
\cite{watanabe2024}). The continuum emission is represented by the \texttt{nlapec} model
\citep{smith2001} in \texttt{xspec}, which is equivalent to the \texttt{apec} model for
thermal plasma emission but without emission lines having emissivities larger than
$10^{-20}$~cm$^3$ s$^{-1}$. Unlike the simple Bremsstrahlung model, \texttt{nlapec} also
includes contributions by the radiative recombination continuum as well as the weak
emission lines.

The model for the Fe Ly$\alpha$ complex was constructed in a similar way. Along with the
continuum emission represented by the \texttt{nlapec} model, four emission lines are
included, representing the two main lines of Fe \emissiontype{XXVI} Ly$\alpha1$ and
Ly$\alpha2$ as well as two satellite lines of Fe \emissiontype{XXV} $J$ and $T$.

The model for the He$\beta$ complex needs to account for contamination from the
overlapping Ni He$\alpha$ line complex. Instead of using \texttt{nlapec} model, we
employed the \texttt{apec} model but removed the three Fe lines in the complex, so that
the model represents both the continuum and the Ni He$\alpha$ lines. Upon it, we
included two main lines of Fe \emissiontype{XXV} He$\beta$ $w$, $y$ ($w_3$, $y_3$
hereafter) and one satellite line of Fe \emissiontype{XXIV} $j_3$.

\paragraph{Fitting}
We performed the fitting using the models described above for each line complex. The
common settings are as follows. Each line component is represented by a Gaussian model
as the Lorentz wing is invisible for the given statistics \citep{aharonian2018}. The
expected width is $\sim$ 0.3 eV for $w$ in the worst-case scenario, which is relatively minor.
The Gaussian line widths are linked to one another. The line centers are fixed to the
reference values (table~\ref{t01}), and the energy shift is fitted collectively. The
parameters of the continuum emission component are initially fixed to those obtained in
the 1.7--10 keV band fitting (\S~\ref{s3-2-1}), with subsequent adjustments made through
renormalization for each complex. The free parameters include the normalization of each
line component, as well as the energy shift and line width of each line complex.

\begin{figure*}[!hbtp]
 \begin{center}
 \includegraphics[width=0.99\linewidth]{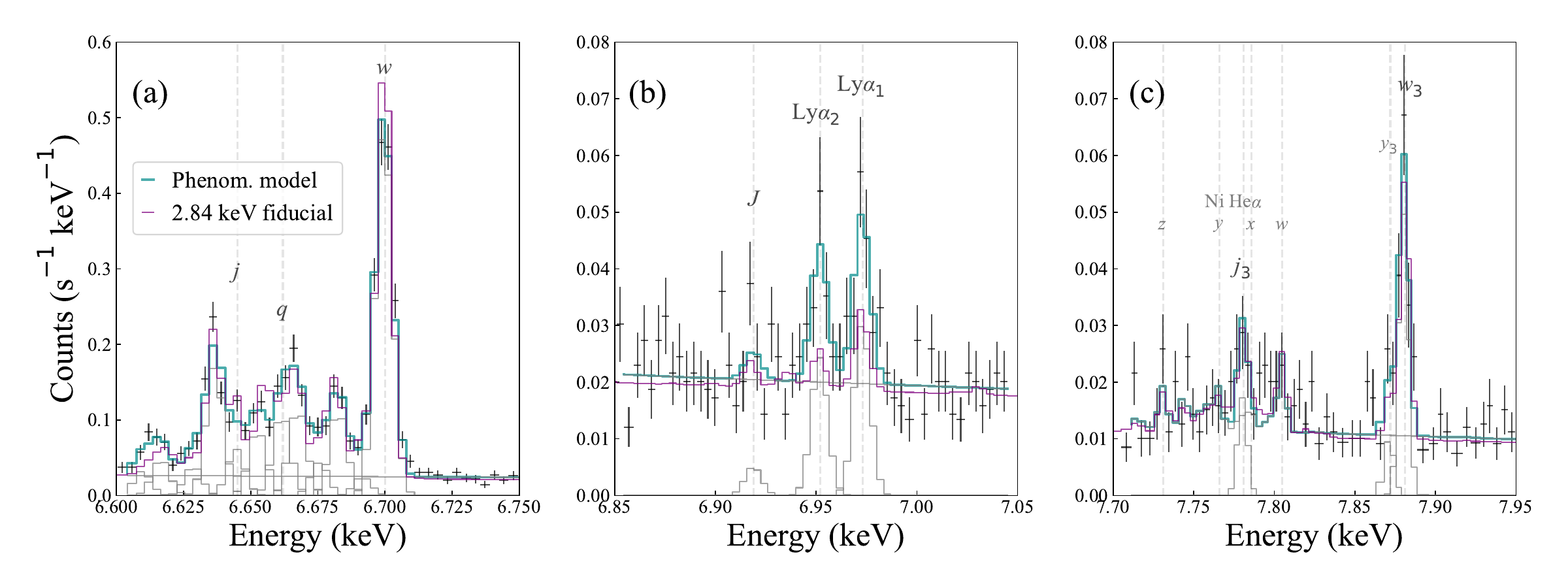}
 \end{center}
 \caption{Close-up views of the \textit{Resolve} spectrum and the best-fit
 phenomenological model for the Fe line complexes: (a) He$\alpha$, (b) Ly$\alpha$, and
 (c) He$\beta$, shown in cyan. The result of the fiducial 2.84~keV thermal plasma model
 (figure \ref{f02}) is also shown in purple. The lines used in the line ratio analysis
 as well as contaminated Ni He$\alpha$ are annotated using names in table~\ref{t01}.
 Alt text: A three-panel figure showing close-up views of the Resolve spectrum and the
 best-fit phenomenological model for Fe line complexes.
 }
 \label{f06}
\end{figure*}

\begin{figure*}[!hbtp]
 \begin{center}
 \includegraphics[width=0.99\linewidth]{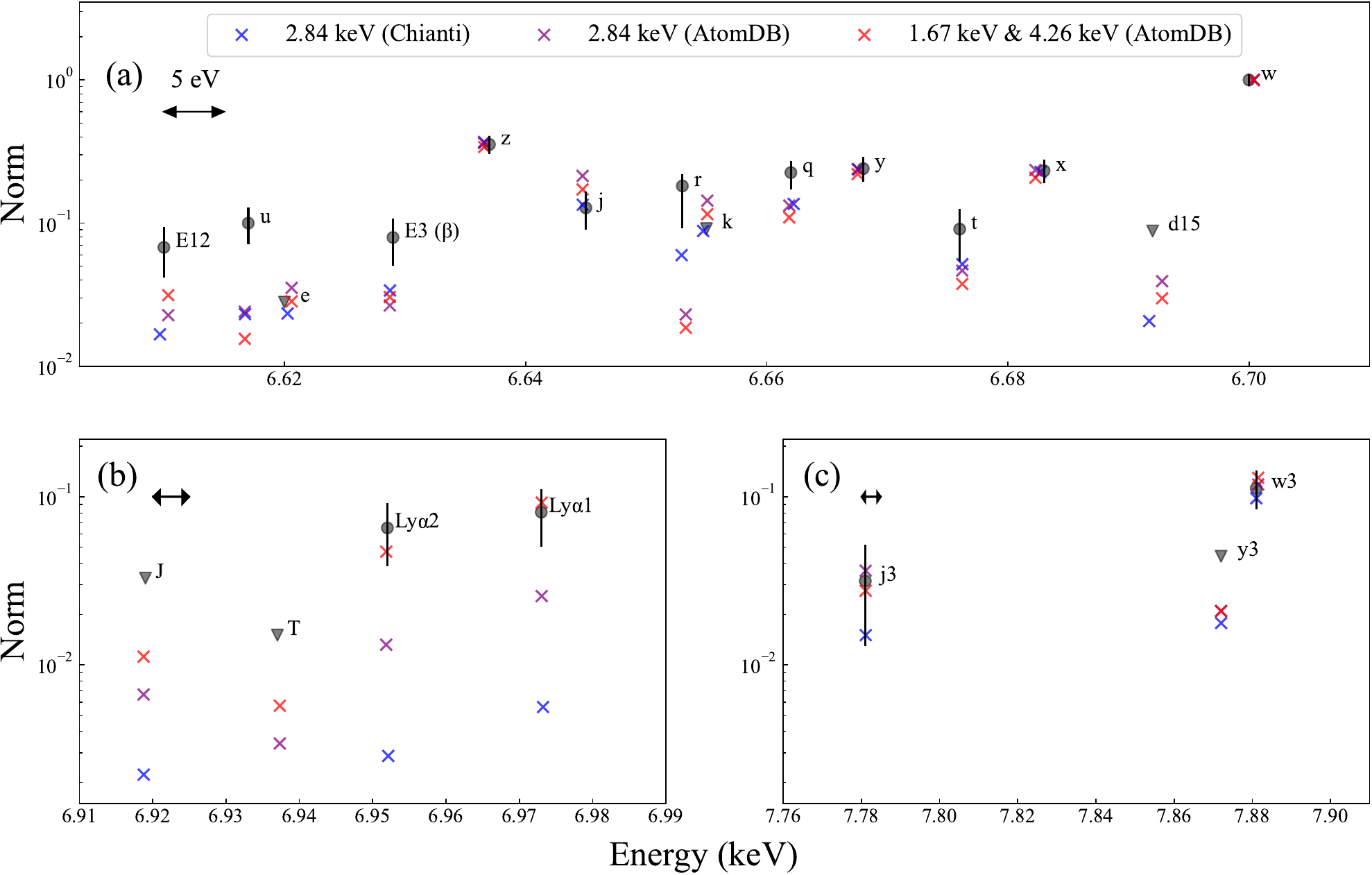}
 \end{center}
 \caption{Comparison of the line strengths between the observation and models for (a) Fe
 He$\alpha$, (b) Ly$\alpha$, and (c) He $\beta$ complexes. The values are normalized to
 He$\alpha$ $w$. The data points are represented by circles with error bars if detected,
 or by triangles if undetected.
 Alt text: A three-panel figures showing the Fe line intensities for each complex. 
 } 
 \label{f07}
\end{figure*}

The best-fit model and the data are illustrated in figure~\ref{f06}, with the best-fit
normalization values shown in figure \ref{f07}. All fits yielded satisfactory results
with the reduced $\chi^2$ values of 43.07/48 for Ly$\alpha$, 50.11/55 for He$\alpha$,
and 46.33/58 for He$\beta$. The line shifts and widths are ($0.06_{-0.03}^{+0.04}$,
$2.4_{-0.29}^{+0.30}$) eV for Fe He$\alpha$, ($0.04_{-0.04}^{+0.19}$,
$3.43_{-3.43}^{+2.12}$) eV for Ly$\alpha$, and ($0.1_{-0.90}^{+0.10}$,
$1.98_{-1.98}^{+2.39}$) eV for He$\beta$. The measured line shifts are consistent with one another but are small compared with the absolute uncertainty on the energy scale. 
For the line width of the He $\alpha$ complex, the systematic uncertainty is relatively small, and the measured width corresponds to 91~km~s$^{-1}$.  If the broadening is entirely attributed to thermal
motion, the width implies an ion temperature of 2.1--4.8 keV. 

\begin{table}
  \tbl{Line list for the phenomenological fitting. }{%
\begin{tabular}{cclclc}
\toprule
Label\footnotemark[$*$] & Ion & Lower& &Upper & $E$ (keV)\footnotemark[$\dag$] \\
\midrule
$E12$  &  Fe \emissiontype{XXIII}& $ \mathrm{1s}^2.\mathrm{2s}.\mathrm{2p}  $~$^3$P$_{2}$ &-&  $\mathrm{1s}.\mathrm{2s}.\mathrm{2p}^2               $~$^3$D$_{3}  $ & 6.6097 \\
$u$\footnotemark[$\ddag$]    &  Fe \emissiontype{XXIV} & $\mathrm{1s}^2.\mathrm{2s} $~$^2$S$_{1/2}$ &-&  $\mathrm{1s}.\mathrm{2s}     ($$^3$S$).\mathrm{2p}   $~$^4$P$_{3/2}  $ & 6.6167 \\
$e$\footnotemark[$\ddag$]    &  Fe \emissiontype{XXIV} & $\mathrm{1s}^2.\mathrm{2p} $~$^2$P$_{3/2}$ &-&  $\mathrm{1s}.\mathrm{2p}^2   ($$^3$P$)      $~$^4$P$_{5/2}  $ & 6.6203 \\
$E3$ ($\beta$) &  Fe \emissiontype{XXIII}& $   \mathrm{1s}^2.\mathrm{2s}^2 $~$^1$S$_{0}$ &-&  $\mathrm{1s}.\mathrm{2s}^2.\mathrm{2p}        $~       $^1$P$_{1}  $ & 6.6288 \\
$ z$    &  Fe \emissiontype{XXV}  & $   \mathrm{1s}^2 $~$^1$S$_{0}$ &-&  $\mathrm{1s}.\mathrm{2s}             $~       $^3$S$_{1}  $ & 6.6366 \\
$j $   &  Fe \emissiontype{XXIV} & $\mathrm{1s}^2.\mathrm{2p} $~$^2$P$_{3/2}$ &-&  $\mathrm{1s}.\mathrm{2p}^2 ($$^1$D$)      $~  $^2$D$_{5/2}  $ & 6.6447 \\
$r $   &  Fe \emissiontype{XXIV} & $\mathrm{1s}^2.\mathrm{2s} $~$^2$S$_{1/2}$ &-&  $\mathrm{1s}.\mathrm{2s}   ($$^1$S$).\mathrm{2p}     $~$^2$P$_{1/2}  $ & 6.6529 \\
$k $   &  Fe \emissiontype{XXIV} & $\mathrm{1s}^2.\mathrm{2p} $~$^2$P$_{1/2}$ &-&  $\mathrm{1s}.\mathrm{2p}^2  ($$^1$D$)       $~$^2$D$_{3/2}  $ & 6.6547 \\
$a $   &  Fe \emissiontype{XXIV} & $\mathrm{1s}^2.\mathrm{2p} $~$^2$P$_{3/2}$ &-&  $\mathrm{1s}.\mathrm{2p}^2  ($$^3$P$)       $~$^2$P$_{3/2}  $ & 6.6579 \\
$q $   &  Fe \emissiontype{XXIV} & $\mathrm{1s}^2.\mathrm{2s} $~$^2$S$_{1/2}$ &-&  $\mathrm{1s}.\mathrm{2s}   ($$^3$S$).\mathrm{2p}     $~$^2$P$_{3/2}  $ & 6.6622 \\
$ y$    &  Fe \emissiontype{XXV}  & $   \mathrm{1s}^2 $~$^1$S$_{0}$ &-&  $\mathrm{1s}.\mathrm{2p}             $~       $^3$P$_{1}  $ & 6.6676 \\
$t$\footnotemark[$\ddag$]     &  Fe \emissiontype{XXIV} & $\mathrm{1s}^2.\mathrm{2s} $~$^2$S$_{1/2}$ &-&  $\mathrm{1s}.\mathrm{2s} ($$^3$S$).\mathrm{2p}     $~  $^2$P$_{1/2}  $ & 6.6762 \\
$x  $   &  Fe \emissiontype{XXV}  & $   \mathrm{1s}^2 $~$^1$S$_{0}$ &-&  $\mathrm{1s}.\mathrm{2p}             $~       $^3$P$_{2}  $ & 6.6827 \\
$d13$   &  Fe \emissiontype{XXIV} & $\mathrm{1s}^2.\mathrm{3p} $~$^2$P$_{3/2}$ &-&  $\mathrm{1s}.\mathrm{2p} ($$^1$P$).\mathrm{3p}     $~  $^2$D$_{5/2}  $ & 6.6892 \\
$d15$   &  Fe \emissiontype{XXIV} & $\mathrm{1s}^2.\mathrm{3p} $~$^2$P$_{1/2}$ &-&  $\mathrm{1s}.\mathrm{2p} ($$^1$P$).\mathrm{3p}     $~  $^2$D$_{3/2}  $ & 6.6917 \\
$ w $    &  Fe \emissiontype{XXV}  & $   \mathrm{1s}^2 $~$^1$S$_{0}$ &-&  $\mathrm{1s}.\mathrm{2p}             $~       $^1$P$_{1}  $ & 6.7004 \\
\midrule
$J$      &  Fe \emissiontype{XXV}  &  $  \mathrm{1s}.\mathrm{2p}$~$^1$P$_{1}$ &-&  $\mathrm{2p}^2  $~$^1$D$_{2}$ &             6.9188 \\
$T$      &  Fe \emissiontype{XXV}  &  $  \mathrm{1s}.\mathrm{2s}$~$^1$S$_{0}$ &-&  $\mathrm{2s}.\mathrm{2p}$~$^1$P$_{1}$ &             6.9373 \\
Ly$\alpha_2$ &  Fe \emissiontype{XXVI} &  $     \mathrm{1s}$~$^2$S$_{1/2}$ &-&  $\mathrm{2p}   $~$^2$P$_{1/2}$ &             6.9521 \\
Ly$\alpha_1$ &  Fe \emissiontype{XXVI} &  $     \mathrm{1s}$~$^2$S$_{1/2}$ &-&  $\mathrm{2p}   $~$^2$P$_{3/2}$ &             6.9732 \\
\midrule
$j_3$  &  Fe \emissiontype{XXIV} & $\mathrm{1s}^2.\mathrm{2p}$~$^2$P$_{3/2}$ &-& $\mathrm{1s}.\mathrm{2p}($$^3$P).3p $^2$D$_{5/2}$    &         7.7811 \\
$y_3$ &  Fe \emissiontype{XXV} & $   \mathrm{1s}^2$~$^1$S$_{0}$ &-& $\mathrm{1s}.\mathrm{3p} $ $^3$P$_{1}$            &           7.8720 \\
$w_3$ &  Fe \emissiontype{XXV} & $   \mathrm{1s}^2$~$^1$S$_{0}$ &-& $\mathrm{1s}.\mathrm{3p} $ $^1$P$_{1}$             &          7.8810 \\
\bottomrule
\end{tabular}
}\label{t01}
\begin{tabnote}
\footnotemark[$*$] Notations are from \citet{doschek1981} for Fe \emissiontype{XXIII}
 lines, \citet{gabriel1972} for Fe \emissiontype{XXIV} $n=2\rightarrow1$ lines, and Safronova's for Fe \emissiontype{XXV} satellite lines. We used the
 same notation of He$\alpha$ for He$\beta$ with a suffix $_{3}$.\\
\footnotemark[$\dag$] Retrieved from \texttt{Chianti} v10.1. \\
\footnotemark[$\ddag$] Blending with weaker lines indecated by \citet{beiersdorfer1993}: $u$ with Fe \emissiontype{XXIV} $v$, Fe \emissiontype{XXIII} $E8$, and Fe \emissiontype{XXIII} $E9$. $e$ with Fe \emissiontype{XXIII} $E6$ and Fe \emissiontype{XXIII} $E7$. $t$ with Fe \emissiontype{XXIV} $m$.    \\ 
\end{tabnote}
\end{table}

\section{Discussion}\label{s4}
In this section, we present the line ratio diagnostics and the interpretation of the
results. We start with a brief description of the physical background (\S~\ref{s4-1})
and apply the diagnostics to the data (\S~\ref{s4-2}).

\subsection{Line ratio diagnostics}\label{s4-1}
\subsubsection{Physics}\label{s4-1-1}
The upper levels of the transitions are populated through various excitation processes,
but occasionally a specific process dominates. Such lines are particularly useful for
line ratio diagnostics due to their straightforward dependence on plasma parameters. In
collisionally-ionized plasmas, two excitation mechanisms play an important role
\citep{phillips2012}: direct excitation (DE) and dielectronic recombination
(DR). Following previous solar observations (e.g., \cite{watanabe2024}), we utilize
their line ratios to characterize the ion charge population and the electron energy
distributions, providing insights into the plasma parameters.

\paragraph{Probing ion charge population}
The ratio of two DE lines from ions of adjacent charges serves as a useful measure to
determine the charge population of the plasma. The DE process, also referred to as the
electron impact excitation process, can be described as
\begin{equation}
 X_i^{+m}+e(E_{\mathrm{kin}}) \quad \rightarrow \quad X_j^{+m,*}+e(E_{\mathrm{kin}}') \quad \rightarrow \quad X_k^{+m}+h \nu (E_{jk}).
\end{equation}
Here, a free electron transfers a part of its kinetic energy $E_{\mathrm{kin}}$ to an
ion, exciting the electron from the state $i$ (mostly the ground state) to $j$. When
they deexcite radiatively to the state $k$ (also mostly the ground state, thus $k=i$) by
emitting a photon, the DE line is formed at an energy $E_{jk}$. The intensity of DE
lines is proportional to the amount of the ion $X^{+m}$. By using lines probing
different $m$ in $X^{+m}$, we can measure the charge population. Because the ionization
and recombination rates depends sensitively on the plasma temperature, the ratio serves
as a reliable diagnostic of the plasma temperature.

\begin{figure}[!hbtp]
 \begin{center}
 \includegraphics[width=0.99\linewidth]{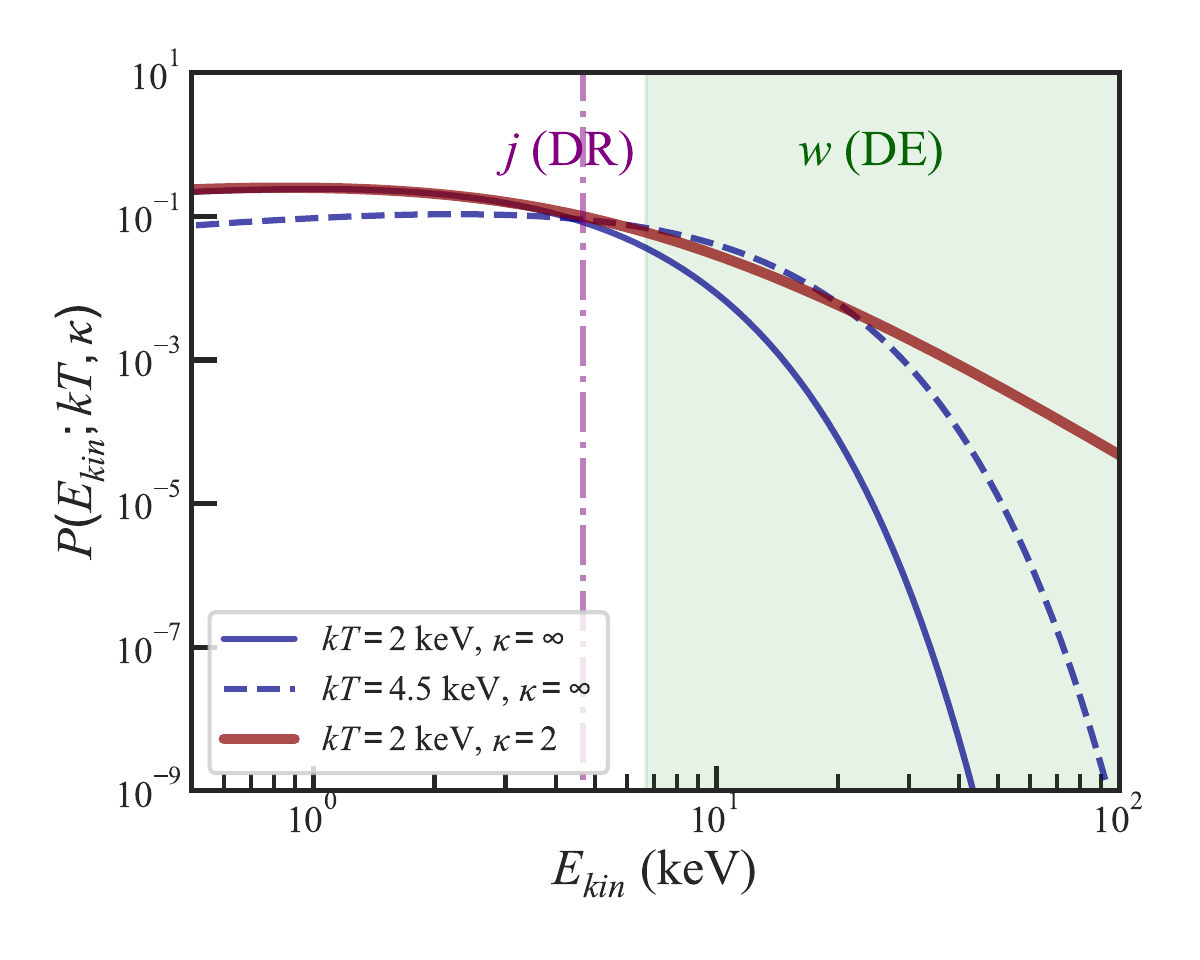}
 \end{center}
 \caption{Three electron energy distributions: the Maxwellian distribution with $kT=$2
 and 4.5~keV, and a $\kappa$ distribution with $kT=$2~keV and $\kappa=2$. The $\kappa$
 distribution is often used to represent a distribution deviated from the Maxwellian
 with a power-law tail characterized by the parameter $\kappa$. It approaches to Maxwell
 distribution as $\kappa\rightarrow\infty$ \citep{oka2013}. The electrons on the purple
 line ($j$) or in the green area ($w$) contribute to the line formation.
  Alt text: A graph
 showing electron energy on the x-axis and electron energy distribution probability on
 the y-axis.
}
\label{f0A}
\end{figure}

\paragraph{Probing electron energy distribution}
For diagnosing electron energy distributions, the different formation mechanisms of DR
and DE lines are leveraged. The DR process can be expressed as
\begin{equation}
 X_i^{+m+1}+e(E_{\mathrm{kin}}) \quad \rightarrow \quad X_j^{+m,**} \quad \rightarrow \quad X_k^{+m}+h \nu (E_{jk})
\end{equation}
Here, when a free electron is captured by the ion $X_i^{+m+1}$, the released energy
(comprising its initial kinetic energy $E_{\mathrm{kin}}$ and the binding energy after
the capture) is used to excite a bound electron, forming a doubly excited state
$X_j^{+m,**}$. This process occurs only for electrons with a specific kinetic energy
that satisfies $E_{\mathrm{kin}}=-I_{\mathrm{pot}}+E_{jk}$ where $I_{\mathrm{pot}}$ is the
ionization potential. As a result, the intensity of the satellite lines produced via DR
reflects the electron population at this specific energy. In contrast, DE lines can
be formed by any electron with kinetic energy greater than $E_{jk}$. This distinction
makes the DR to DE line ratio a valuable probe of the electron energy distribution
\citep{gabriel1972,gabriel1979}. 
Figure \ref{f0A} illustrates this behavior by showing the excitation energies of DE Fe
\emissiontype{XXV} $w$ and DR Fe \emissiontype{XXIV} $j$ along with a comparison of
different electron energy distributions.

For a thermal plasma, the DR to DE ratio can be expressed as a function of the
temperature (see e.g., \cite{phillips2012}):
\begin{equation}
\frac{I_{\mathrm{DR}}}{I_{\mathrm{DE}}} \propto \frac{\exp \left[\left(\Delta E_{\mathrm{DE}}-\Delta E_{\mathrm{DR}}\right) / kT\right]}{kT}
\end{equation}
The transition energy of DE and DR lines ($\Delta E_{\mathrm{DE}}$ and $\Delta
E_{\mathrm{DR}}$) are usually close, thus the ratio is approximately inversely
proportional to temperature, making it a useful diagnostic tool.  Moreover, the ratio is
highly sensitive to deviations from the Maxwellian distribution, such as the $\kappa$
distribution, which exhibits an enhanced high-energy tail (figure~\ref{f0A}). In
addition to its strong dependence on the electron energy distribution, this ratio has
the advantage of being independent of charge state distributions, as both lines are
produced from the same parent ion.

\subsubsection{Pairs in the Fe K band}\label{s4-1-2}
We apply the diagnostics described in \S~\ref{s4-1} to the present data in the Fe K
band. The strongest lines in the three complexes are Fe \emissiontype{XXV} $w$ for
He$\alpha$, Fe \emissiontype{XXVI} Ly$\alpha_1$ for Ly$\alpha$, and $w_3$ for
He$\beta$. All of these lines originate mainly from DE processes and undergo radiative
decay via an electric dipole transition to the ground state.

In the He$\alpha$ complex, many satellite lines are detected (figure~\ref{f06}a). The Fe
\emissiontype{XXIV} $q$ line is populated primarily by inner-shell excitation from
the ground state of Li-like Fe \citep{bely1982}. The Fe \emissiontype{XXIV} $j$ line, on
the other hand, is a DR line that reflects the population of He-like Fe
\citep{bely1982}. Another well-known DR line, Fe \emissiontype{XXIV} $k$ (e.g.,
\cite{beiersdorfer1992,phillips2012}), is significantly blended with the Fe
\emissiontype{XXIV} $r$ line in our dataset. These DR satellite lines are formed
with a spectator electron in the L shell, but a similar DR process can also occur with a
spectator electron in the M shell or higher. Some of these lines (e.g, $d13$ and $d15$
lines) were included in our phenomenological fitting; however, they are blended with the
Fe \emissiontype{XXV} $w$ line, making their individual contributions difficult to
isolate.
In the Ly$\alpha$ complex (figure~\ref{f06}b), only upper limits were obtained for the
two satellite lines, Fe \emissiontype{XXV} $J$ and $T$. Among them, the $J$ line was
better constrained and thus used in our analysis.  In the He$\beta$ complex
(figure~\ref{f06}c), $j_3$ is the only DR line clearly detected, which corresponds to a
$\mathrm{1s}-\mathrm{3p}$ equivalent of the $\mathrm{1s}-\mathrm{2p}$ transition
associated with $j$ line in the He$\alpha$ complex.

Among these lines, we used two line ratios to investigate the ion charge distribution and
three line ratios to probe the electron energy distribution. The former consists of (i)
Fe \emissiontype{XXIV} $q$ / Fe \emissiontype{XXV} $w$ and (ii) Fe \emissiontype{XXVI}
Ly$\alpha_1$ /Fe \emissiontype{XXV} $w$. The latter consists of (iii) Fe
\emissiontype{XXIV} $j$ / Fe \emissiontype{XXV} $w$ in the He$\alpha$ complex, (iv) Fe
\emissiontype{XXIV} $j_3$ / Fe \emissiontype{XXV} $w_3$ in the He$\beta$ complex, and
(v) Fe \emissiontype{XXV} $J$ and Fe \emissiontype{XXVI} Ly$\alpha_1$ in the Ly$\alpha$
complex. In each case, the stronger of the two lines was used as the denominator in the ratio.

For comparison, we used two spectral synthesis codes for the thermal plasma; one is
\texttt{apec}/\texttt{AtomDB} v3.0.9 \citep{smith2001,foster2012} and the other is
\texttt{Chianti} v10.1 \citep{dere1997,dere2023}. The former is traditionally used in
astrophysical plasmas, while the latter is more common in solar plasmas. They have their
own atomic databases of different origins. Their differences were studied in many use
cases, including the analyses using the SXS data \citep{aharonian2018}. For some Fe
K-shell lines considered here, we found that the discrepancy can be as large as a factor
two. We do not aim to determine which database is more accurate in this paper. Instead,
we present their differences to highlight the need for further refinements in atomic
data.

\subsection{Application}\label{s4-2}
\subsubsection{Single-temperature plasma}\label{s4-2-1}
\begin{figure}[!hbtp]
 \begin{center}
 \includegraphics[width=0.99\linewidth]{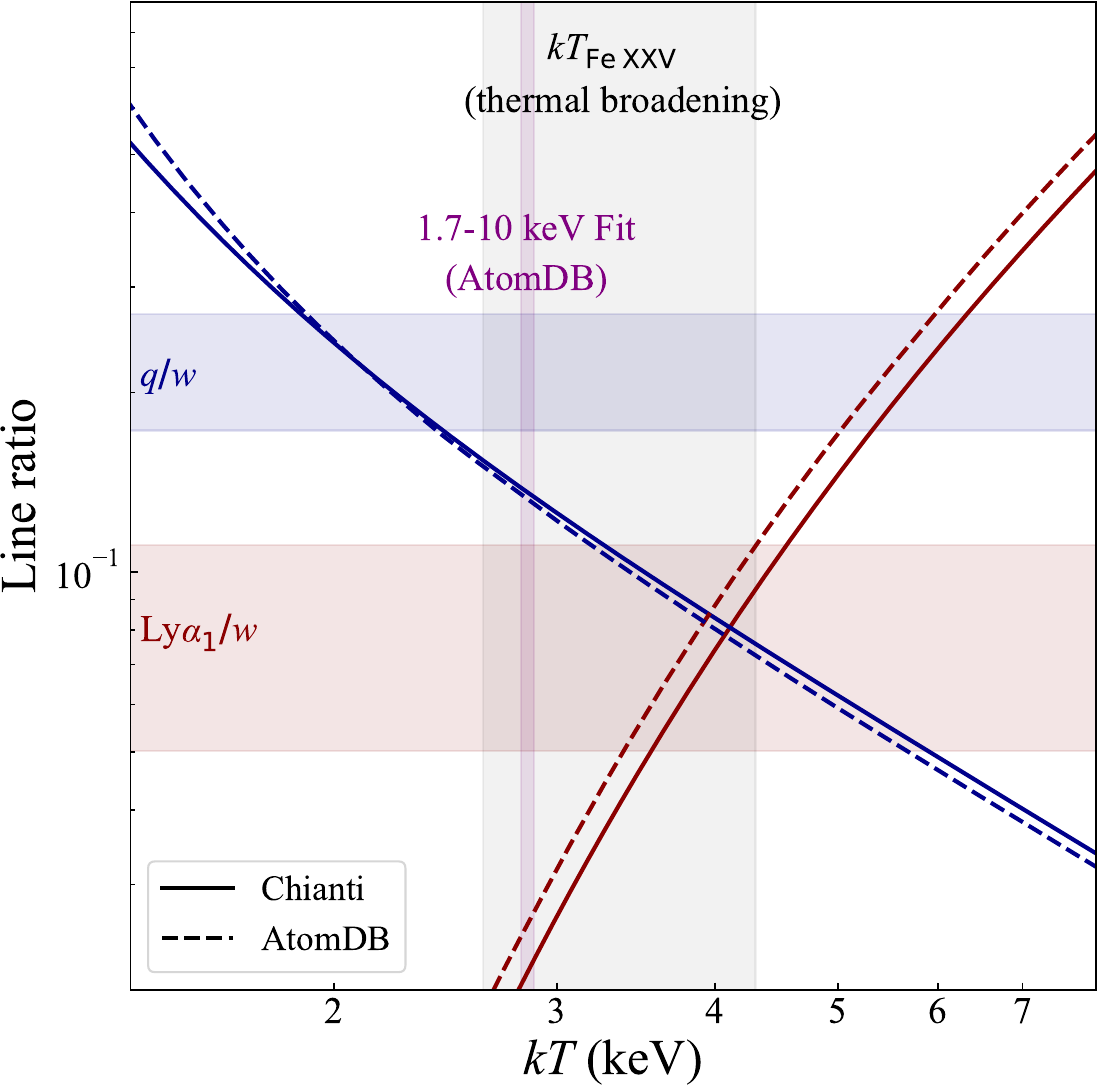}
 \end{center}
 \caption{Temperature diagnostics using line ratios sensitive to ion charge
 distribution. The line ratios of $q$/$w$ (Li-like/He-like Fe) and Ly$\alpha_1$/$w$
 (H-like/He-like Fe) are shown in blue and red, respectively, for the observation
 (shaded areas) and calculation using \texttt{Chianti} (solid curves) and
 \texttt{AtomDB} (dashed curves). The best-fit $kT$ from the broadband fitting using
 \texttt{AtomDB} (\S~\ref{s3-2-1}) and $kT_{\mathrm{Fe}\emissiontype{XXV}}$ derived from
 the line broadening in the He$\alpha$ complex are also presented.  Alt text: A graph
 showing temperature on the x-axis and line ratio on the y-axis.  }
 \label{f03}
\end{figure}

\paragraph{Ion charge distribution}
We first examined the charge distribution using $q$/$w$ and Ly$\alpha_1$/$w$ ratios. The
$q$/$w$ ratio probes the relative charge fraction between Li-like and He-like Fe ions,
while the Ly$\alpha_1$/$w$ ratio probes that between He-like and H-like Fe ions. The
expected line ratios are calculated as a function of $kT$ based on \texttt{Chianti} and
\texttt{AtomDB} and compared with the observed ratios (figure \ref{f03}). In this
temperature range, the fraction of more highly ionized species increases monotonically,
causing the $q$/$w$ ratio to decreases and the Ly$\alpha_1$/$w$ ratio to increases as
$kT$ increases.

The observed $q$/$w$ ratio suggests $kT \sim$2~keV, while the Ly$\alpha$/$w$ ratio
indicates $\sim$4~keV. This clearly demonstrates that a single-temperature
representation is inadequate for this plasma. Indeed, fitting the broadband spectrum
with a single-temperature (\S~\ref{s3-2-1}) resulted in a best-fit temperature of
2.84~keV, which is inconsistent with both line ratios.

\paragraph{Electron energy distribution}
\begin{figure}[!hbtp]
 \begin{center}
 \includegraphics[width=0.99\linewidth]{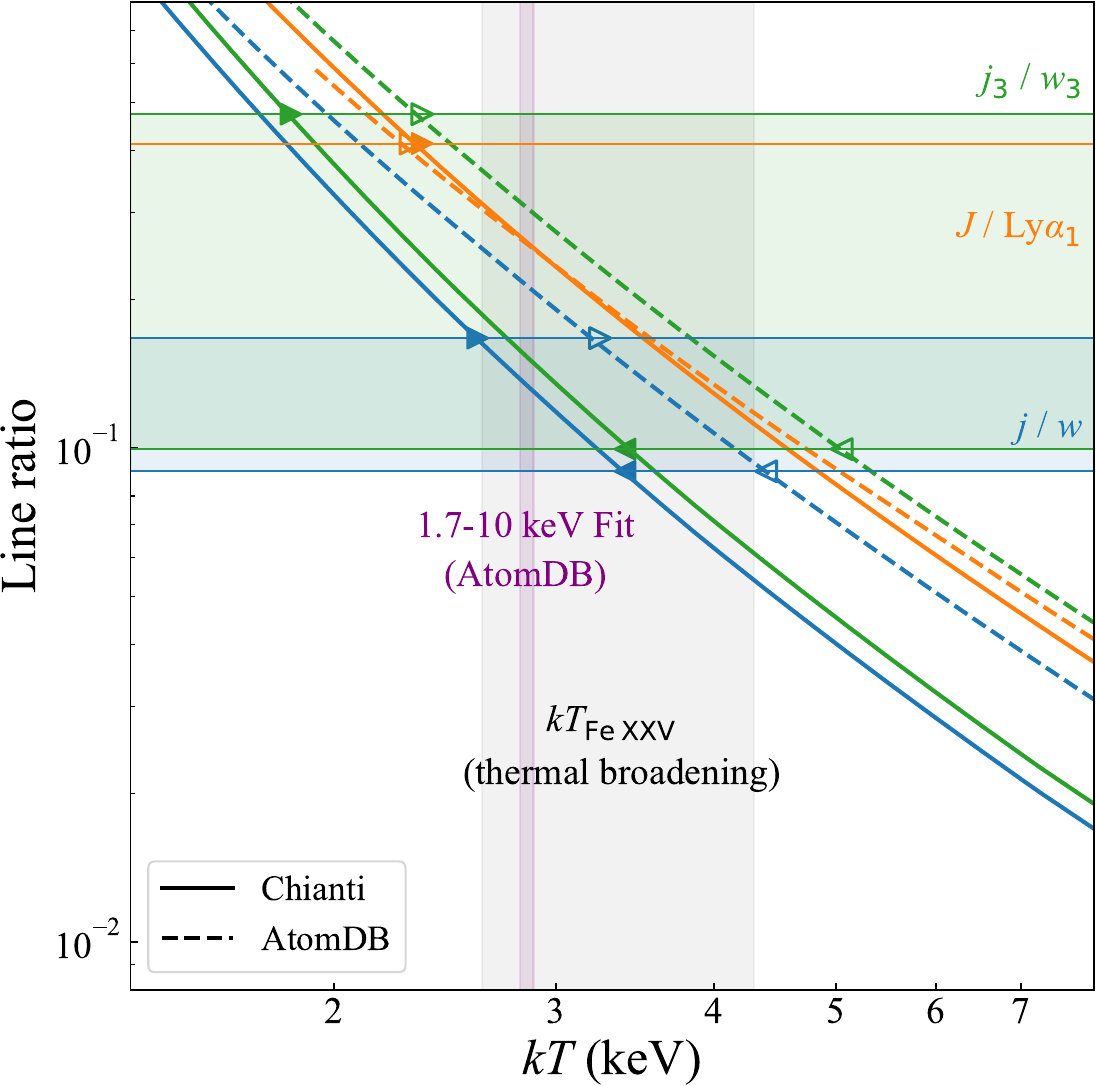}
 \end{center}
 \caption{Temperature diagnostics using line ratios sensitive to electron energy
 distribution. The line ratios of $j$/$w$, $J$/Ly$\alpha_1$, and $j_3$/$w_3$ are shown
 in blue, orange, and green, respectively for the observation (shaded area and upper and
 lower limit lines for $j$/$w$ and $j_3$/$w_3$, while only the upper limit line is shown
 for $J$/Ly$\alpha_1$) and calculations using \texttt{Chianti} (solid curves) and
 \texttt{AtomDB} (dashed curves). The best-fit $kT$ from the broadband fitting using
 \texttt{AtomDB} (\S~\ref{s3-2-1}) and $kT_{\mathrm{Fe}\emissiontype{XXV}}$ derived from
 the line broadening in the He$\alpha$ complex are also presented.  Alt text: A graph
 showing temperature on the x-axis and line ratio on the y-axis. } 
 \label{f04}
\end{figure}

We next performed diagnostics using the three DR to DE line ratios
(figure~\ref{f04}). In this temperature range, as the plasma temperature $kT$ increases,
the population of higher energy electrons in the electron energy distribution increases
(figure~\ref{f0A}), causing the DR/DE line ratio to decrease with increasing $kT$.  The
two databases produce different results, with \texttt{Chianti} systematically predicting
the smaller values for all three ratios compared to \texttt{AtomDB} at a given
$kT$. Using \texttt{Chianti}, we found that $kT=2.6-3.4$~keV satisfies all three
ratios. Using \texttt{AtomDB}, $3.3-4.3$~keV satisfies all the three ratios. This is
again inconsistent with the broadband fitting result using \texttt{AtomDB} of 2.84~keV.

\subsubsection{Deviations from single-temperature plasma}\label{s4-2-2}
We found that a single-temperature representation, which worked for the broadband
fitting (\S~\ref{s3-2-1}), does not work for the line ratio diagnostics
(\S~\ref{s4-2-1}). We now explore three deviations from it below.

\begin{figure*}[!hbtp]
 \begin{center}
 \includegraphics[width=0.99\linewidth]{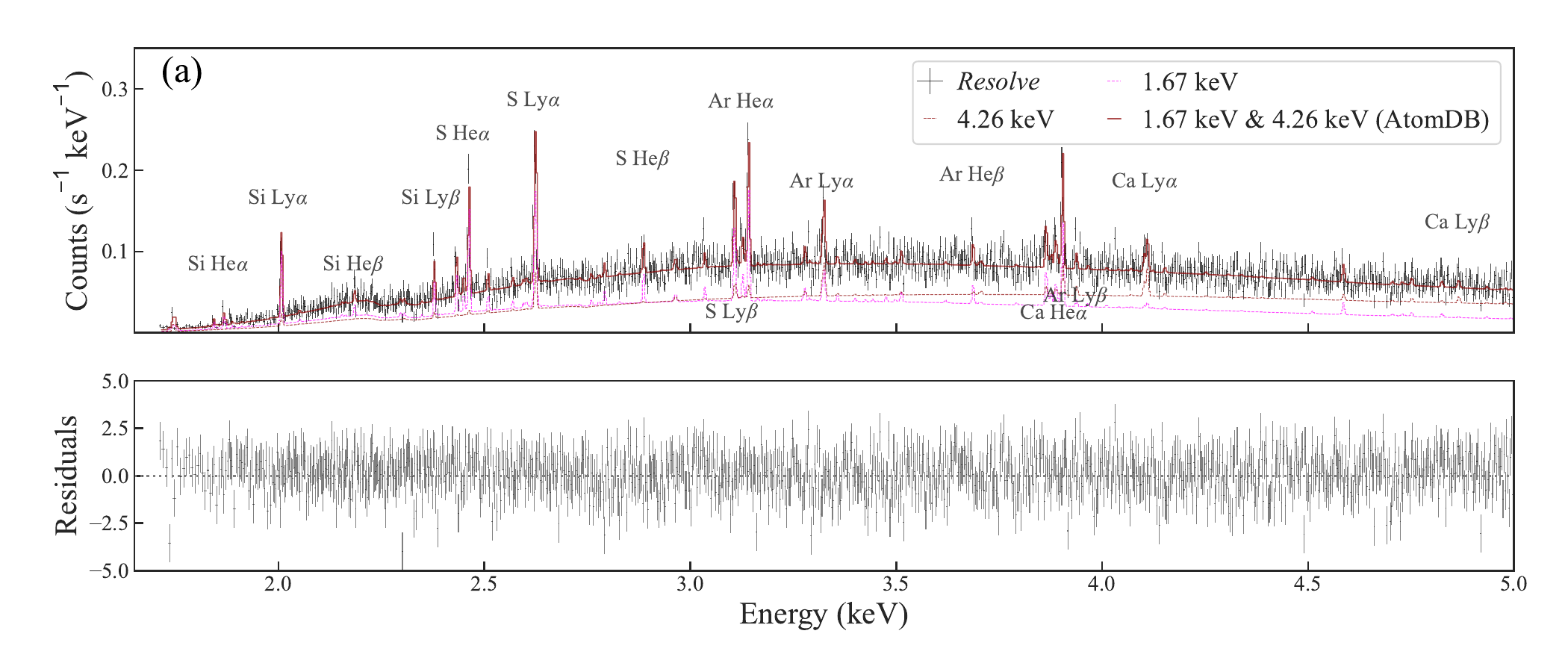}
 \includegraphics[width=0.99\linewidth]{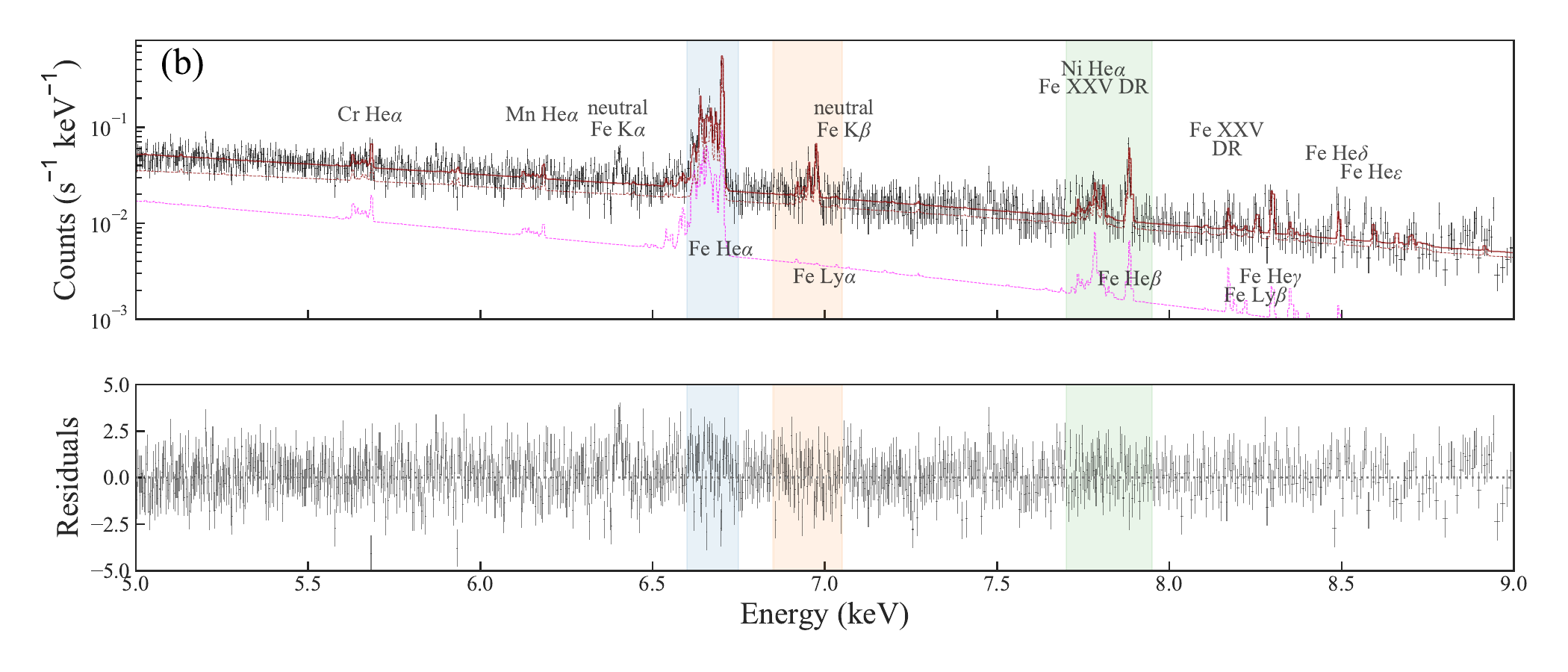}
 \end{center}
 \caption{Two-temperature plasma model and the residuals from the data in the (a)
 1.7--5.0 and (b) 5.0--9.0 keV band.
 Alt text: a couple of two-panel figures showing an X-ray spectrum in the 1.7–10.0 keV range.
 }
 \label{f09}
\end{figure*}

\paragraph{Multi-temperature plasma}\label{s4-2-2-1}
The simplest deviation from a single-temperature model is to introduce another
single-temperature component of a different $kT$. We revisited the broadband spectral
fitting and applied the two-temperature plasma model. The free parameters include the
abundances of major elements and the ISM absorption column ($N_{\mathrm{H}}$) common
between the two components, and the plasma temperature $kT_{i}$ and emission measure
$EM_{i}$ of the two components ($i \in \{1,2\}$). By constraining $kT_1 <2$~keV and
$kT_2>2$~keV, we obtained improved results over the one-temperature fitting with a reduced $\chi^2 = 2203.43 / 2169$. The
best-fit parameters are shown in table~\ref{t02}, while the model is in
figure~\ref{f09}.

\begin{table}[]
  \tbl{Two-temperature thermal plasma model fit results}{%
\begin{tabular}{ccc}
\midrule
Parameters                  &       low-$kT$ component                    &           high-$kT$ component                         \\
\midrule
$N_H$ ($10^{21}$ cm$^{-2}$) & \multicolumn{2}{c}{1.3$^{+1.7}_{-1.3}$} \\
$kT$ (keV)        & $1.67^{+0.10}_{-0.14}$                    & $4.26^{+0.31}_{-0.30}$                   \\
$EM$ ($10^{54}$ cm$^{-3}$)  & $26^{+3.1}_{-3.0}$                 & $13^{+2.6}_{-1.9}$                       \\
$Z_{\mathrm{Si}}$ ($Z_{\odot}$) & \multicolumn{2}{c}{$0.35^{+0.06}_{-0.05}$} \\                                       
$Z_{\mathrm{S}}$ ($Z_{\odot}$) & \multicolumn{2}{c}{$0.32^{+0.03}_{-0.03}$} \\                                       
$Z_{\mathrm{Ar}}$ ($Z_{\odot}$) & \multicolumn{2}{c}{$0.48^{+0.07}_{-0.06}$} \\                         
$Z_{\mathrm{Ca}}$ ($Z_{\odot}$) & \multicolumn{2}{c}{$0.51^{+0.07}_{-0.07}$} \\                              
$Z_{\mathrm{Fe}}$ ($Z_{\odot}$) & \multicolumn{2}{c}{$0.28^{+0.01}_{-0.01}$} \\                                  
$Z_{\mathrm{Ni}}$ ($Z_{\odot}$) & \multicolumn{2}{c}{$0.37^{+0.16}_{-0.16}$} \\                                                 
\midrule
\end{tabular}}\label{t02}
\end{table}

The two-temperature model is also favored when considering the Fe line ratios. In
figure~\ref{f07}, two-temperature model enhances the consistency between the observed
and calculated line strengths, particularly for the Ly$\alpha$ lines. However, we still
observe that the Fe \emissiontype{XXIV} $q$ and $t$ lines and Fe \emissiontype{XXIII}
$E12$ and $E3$ lines are stronger in the observation than in the calculation, which may
suggest the presence of an even lower temperature component \citep{doschek1981}.

\paragraph{Suprathermal electron distribution}
Next, instead of adding another thermal plasma component, we investigated whether
modifying the electron energy distribution could explain the observed line ratios. The
$\kappa$ distribution is commonly used in solar corona (e.g., \cite{dudik2021}) to
represent the distribution with a power-law suprathermal tail upon a thermal Maxwellian
distribution. The parameter $\kappa \in [1.5, \infty)$ represents the relative fraction
of the suprathermal population, while $kT_{\mathrm{M}}$ is for the main thermal
population. At $\kappa \rightarrow \infty$, the distribution asymptotes to the
Maxwellian distribution of a $kT_{\mathrm{M}}$ temperature.

\begin{figure*}[!hbtp]
 \begin{center}
 \includegraphics[width=0.49\linewidth]{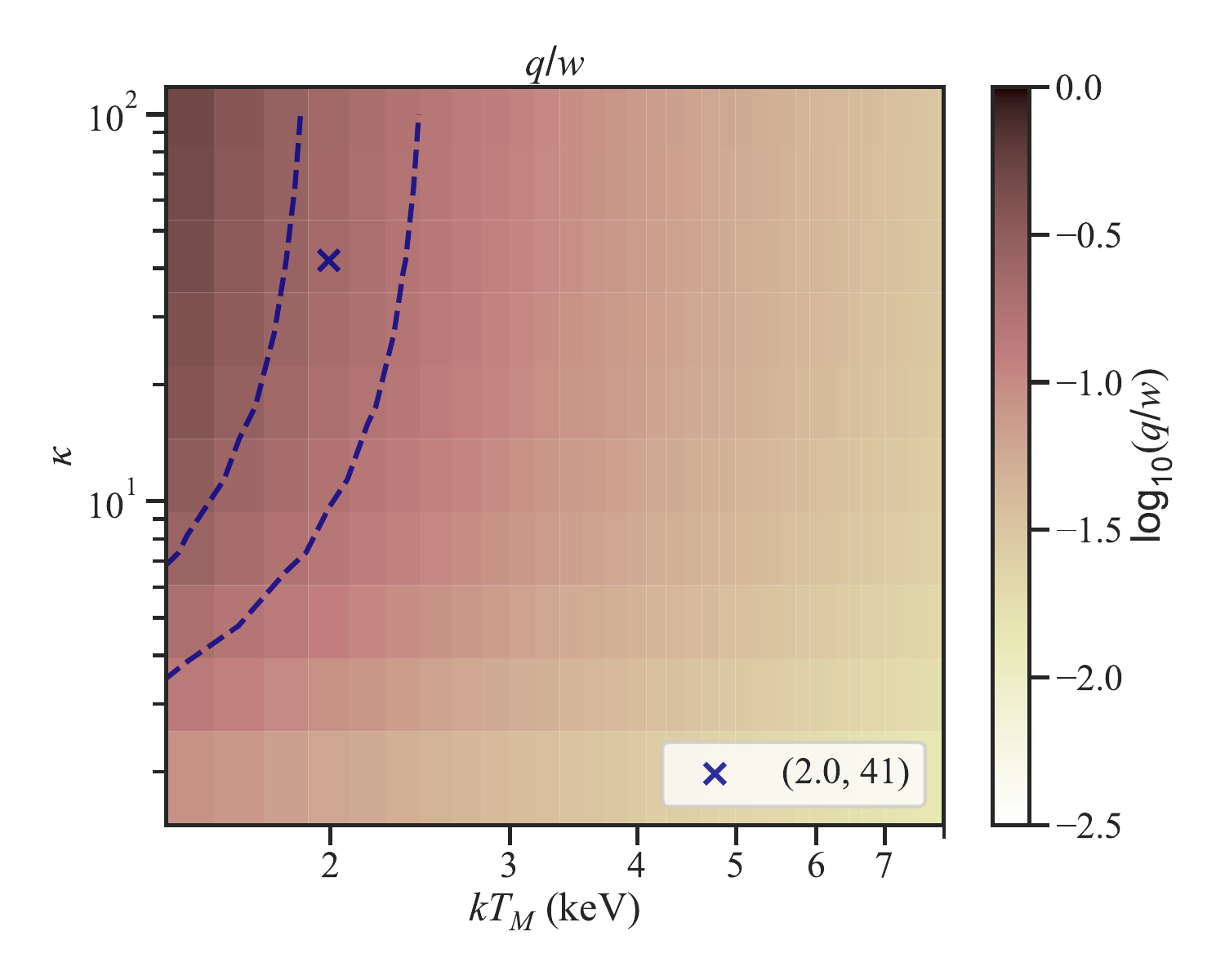}
 \includegraphics[width=0.49\linewidth]{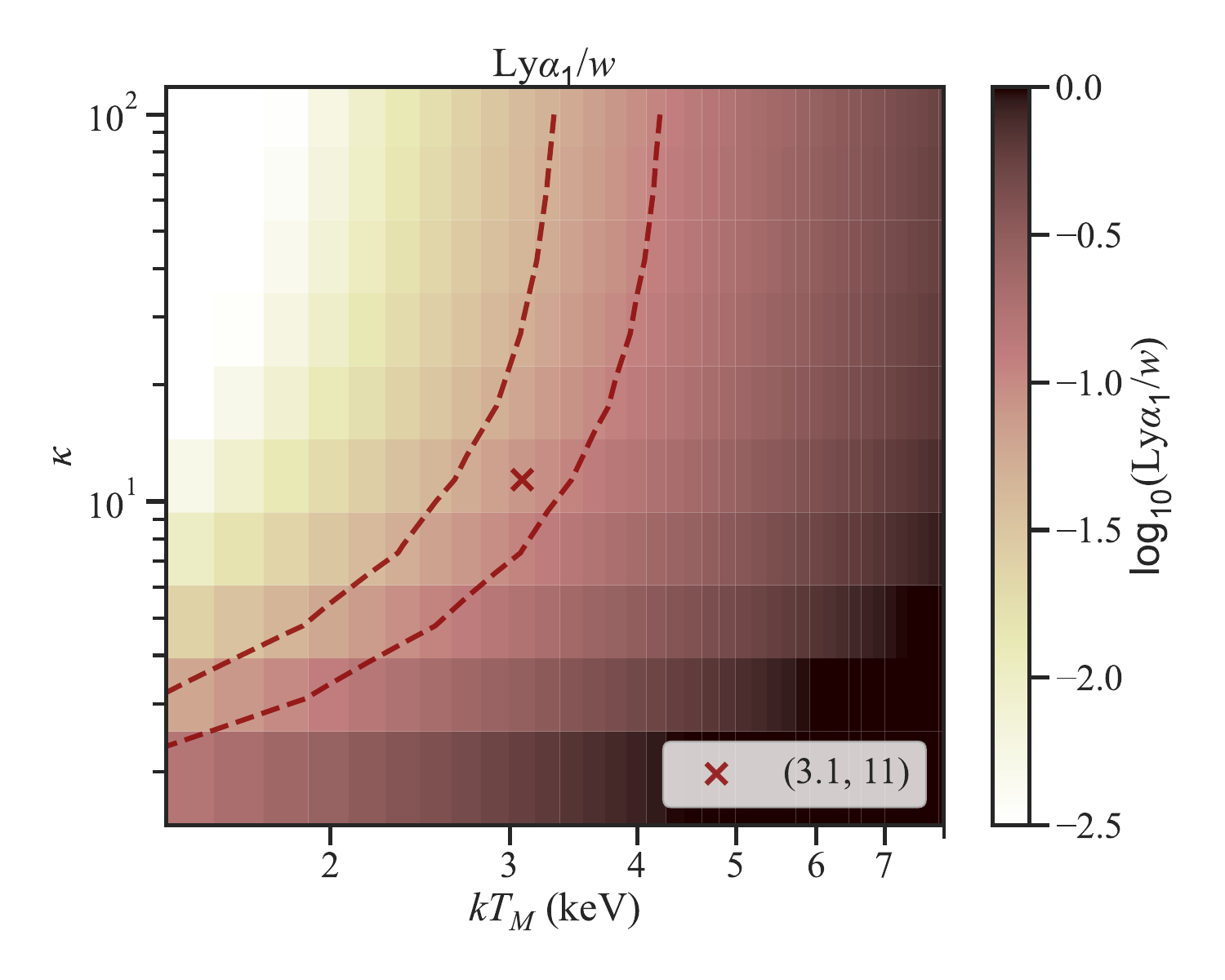}
 \includegraphics[width=0.49\linewidth]{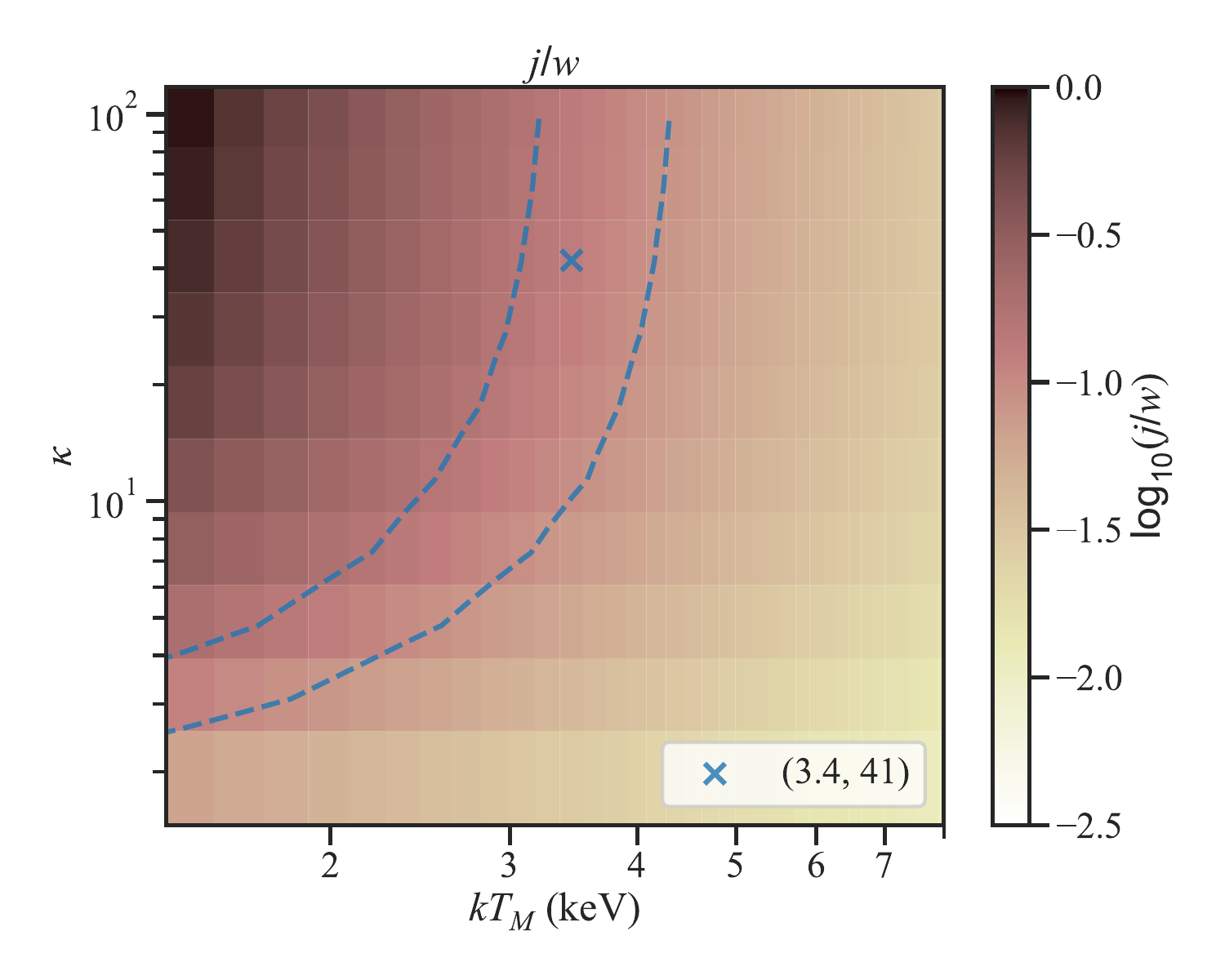}
 \includegraphics[width=0.49\linewidth]{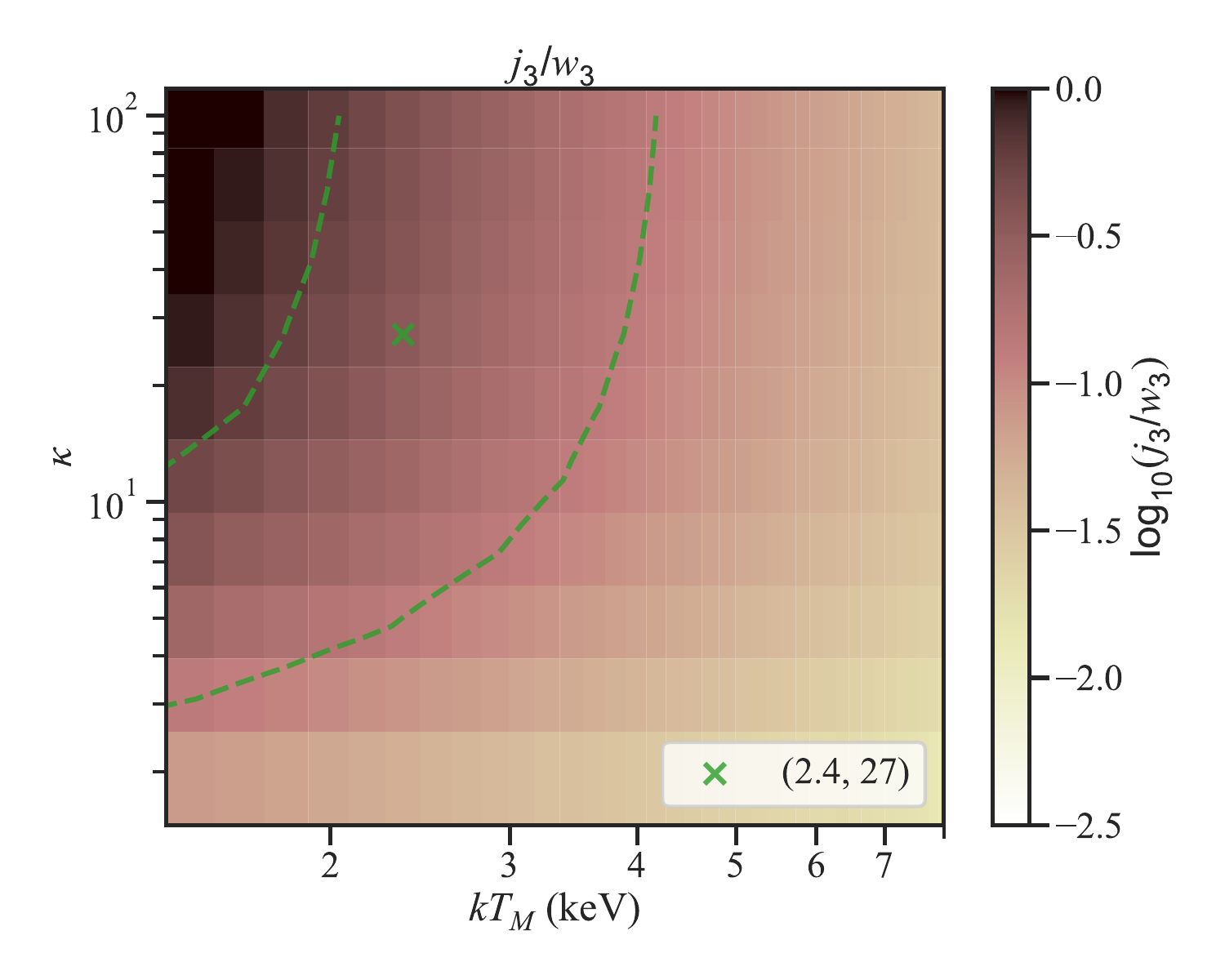}
 \includegraphics[width=0.49\linewidth]{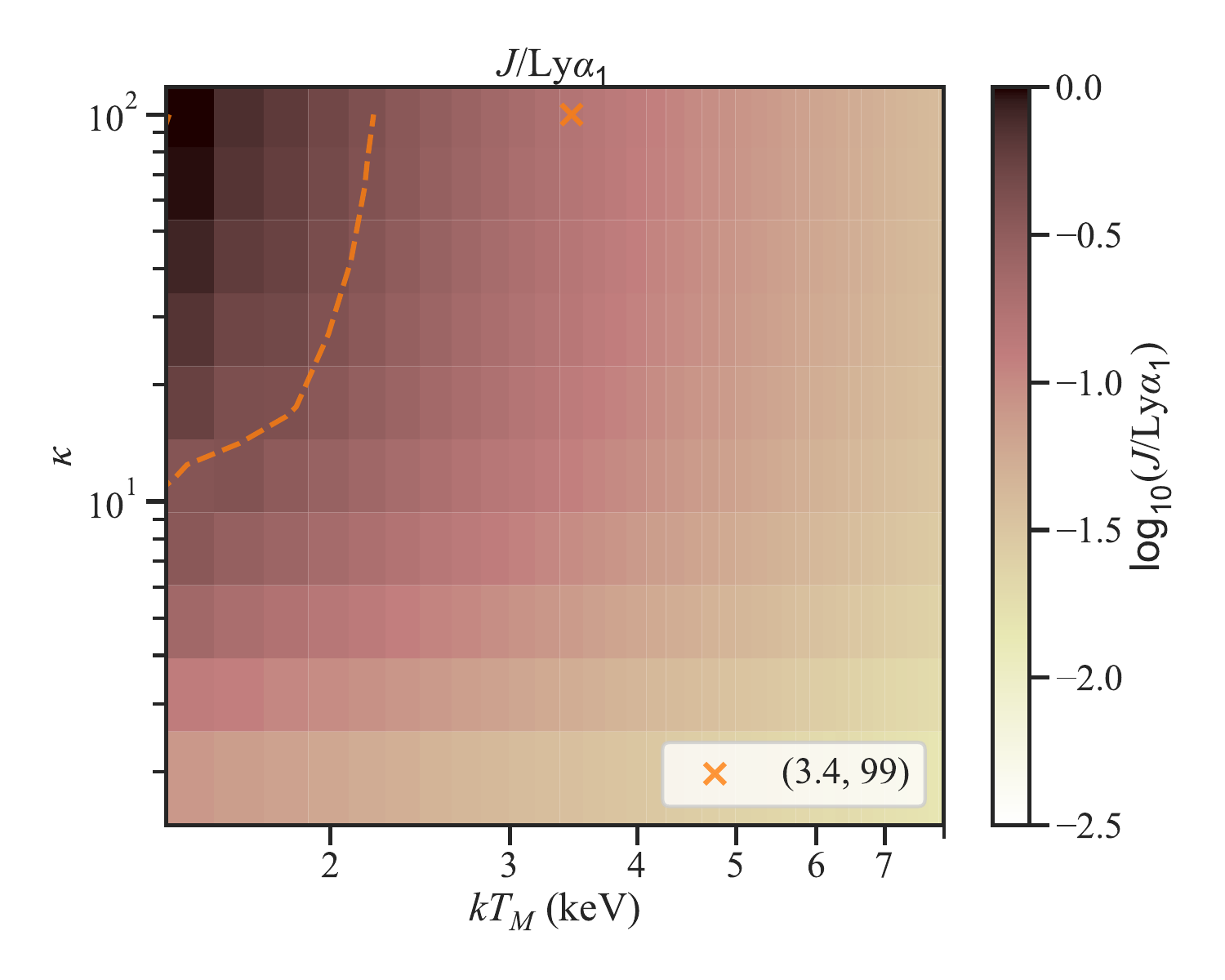}
 \includegraphics[width=0.49\linewidth]{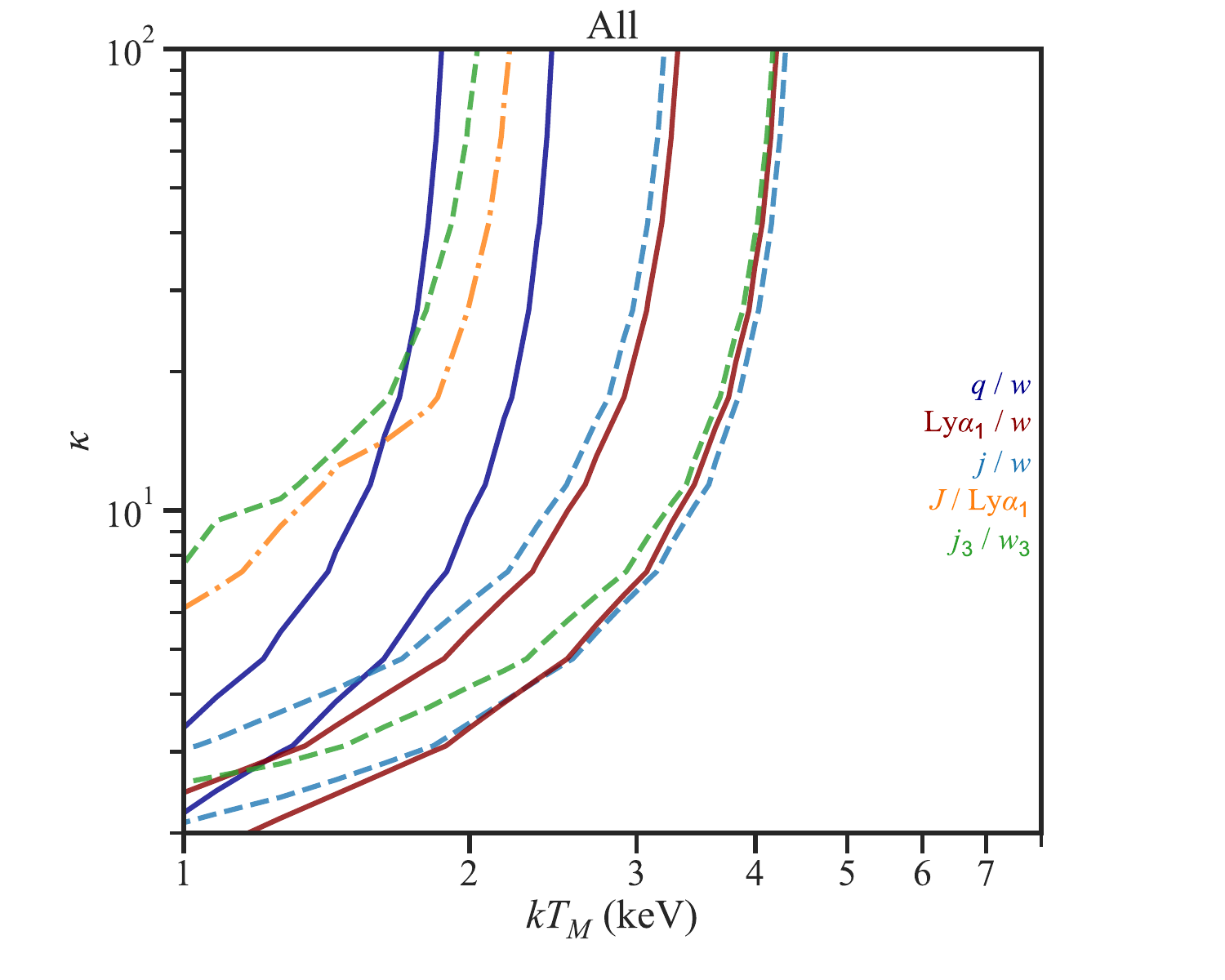}
 \end{center}
 \caption{Diagnostics for the $\kappa$ electron energy distribution. Heatmaps are
 created for five line ratios based on \texttt{AtomDB} model \citep{cui2019}, covering the
 two parameters $\kappa$ and $kT_{\mathrm{M}}$. The best-fit values of the two
 parameters are marked with a cross, and the 90\% uncertainty range is indicated with
 contours. Only the upper limit is provided for the $J$/Ly$\alpha$1 ratio. All contours are
 assembled in the bottom right. The contour line style distinguishes the two pairs for ion
 charge distribution (solid lines) and three pairs for electron energy distribution
 (dashed lines). No space satisfies all the contours simultaneously.
 Alt text: a six-panel figure of heatmaps showing $kT_{\mathrm{M}}$ on the x-axis and $\kappa$ on the y-axis.
 }
 \label{f08_kappa}
\end{figure*}

Using the $\kappa$ model based on \texttt{AtomDB} \citep{cui2019}, we calculated the
five line ratios as a function of $\kappa$ and $kT_{\mathrm{M}}$ and restricted their
values using the observed ratio (figure~\ref{f08_kappa}). No combination of $\kappa$ and
$kT_{\mathrm{M}}$ can explain all the observed ratios. Therefore, we conclude that the
suprathermal electron distribution is not justified in a way that accounts for the
observed line ratios.

\paragraph{Non-equilibrium plasma}
We also investigated the possibility that the electron ($kT_{\mathrm{e}}$) and
ionization ($kT_{\mathrm{Z}}$) temperatures are different in two ways: one is the ionizing
plasma ($kT_{\mathrm{e}} > kT_{\mathrm{Z}}$) and the other is the recombining plasma
($kT_{\mathrm{e}} < kT_{\mathrm{Z}}$). Both scenarios are characterized by two
parameters; the electron temperature $kT_{\mathrm{e}}$ and the Coulomb relaxation time
scale $\tau \equiv \int n_{\mathrm{e}} dt$, in which $n_{e}$ is the electron density.

\begin{figure*}[!hbtp]
 \begin{center}
 \includegraphics[width=0.49\linewidth]{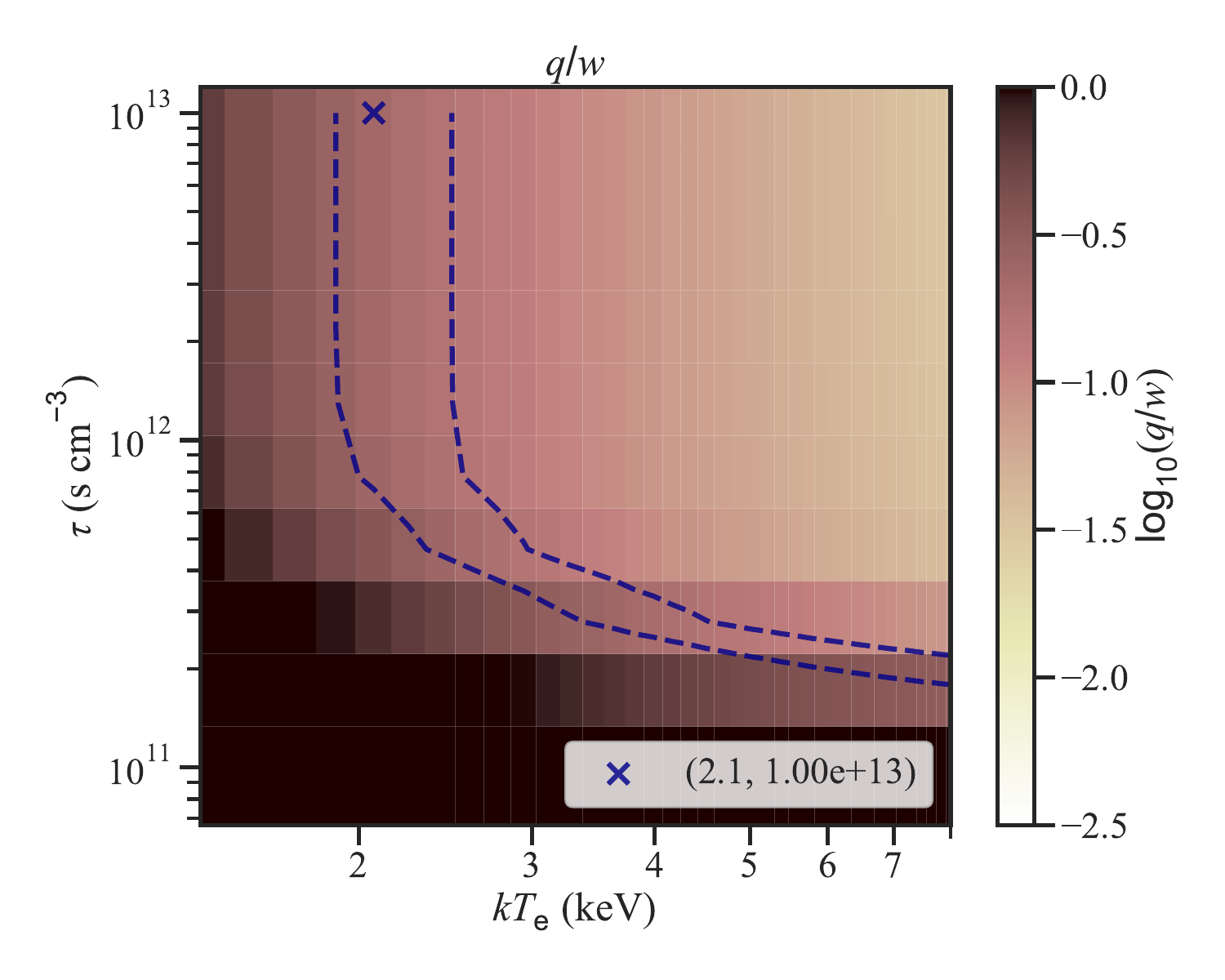}
 \includegraphics[width=0.49\linewidth]{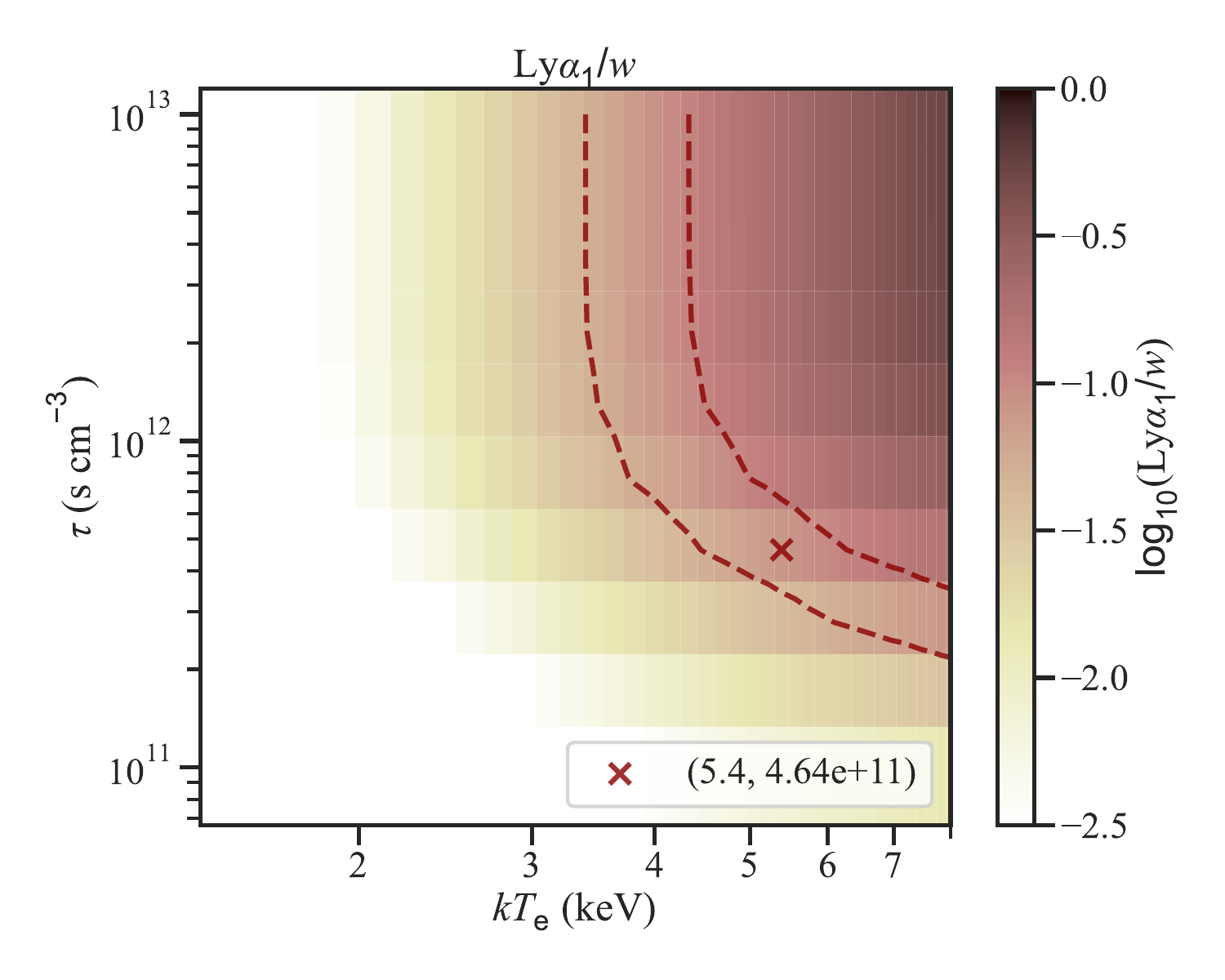}
 \includegraphics[width=0.49\linewidth]{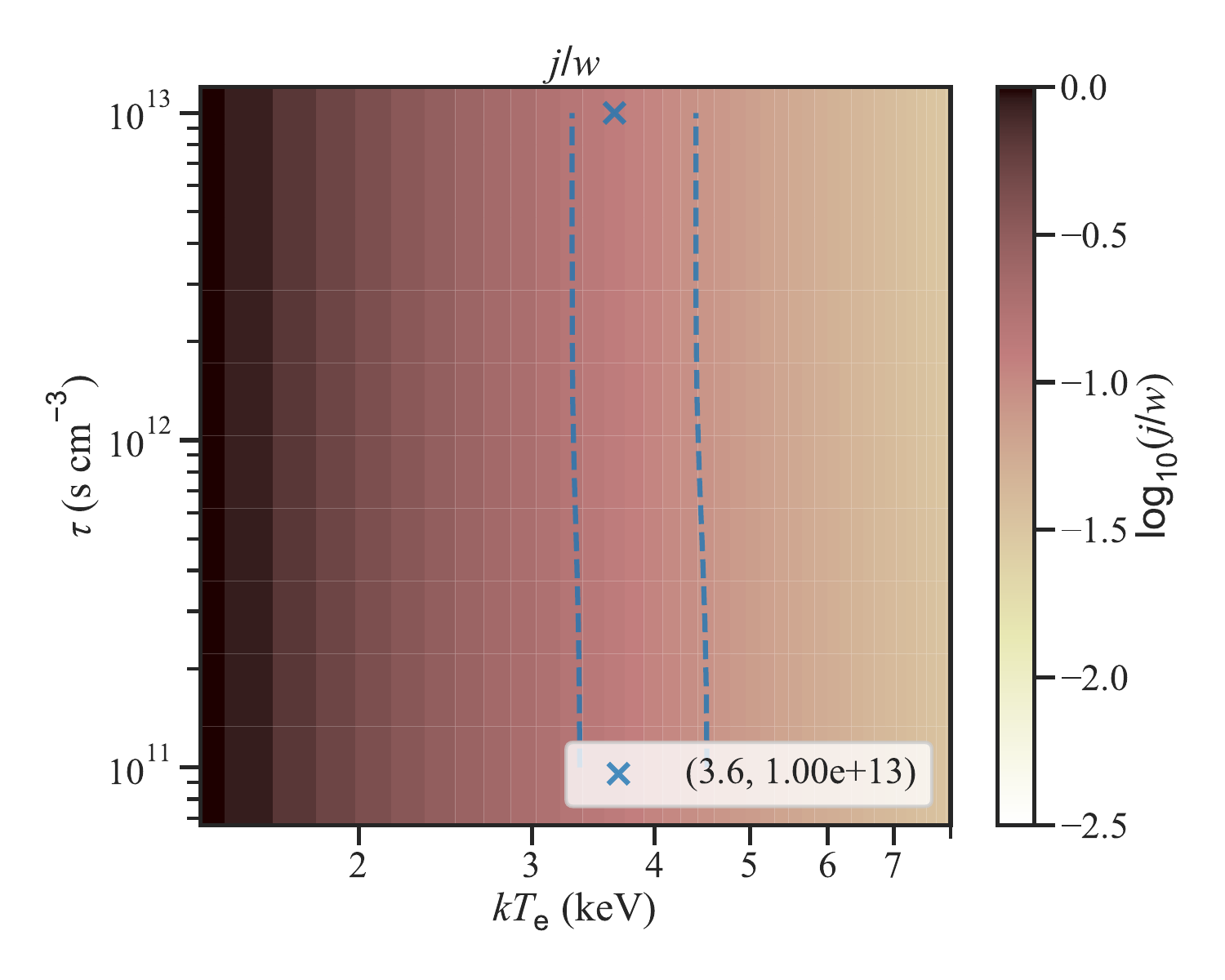}
 \includegraphics[width=0.49\linewidth]{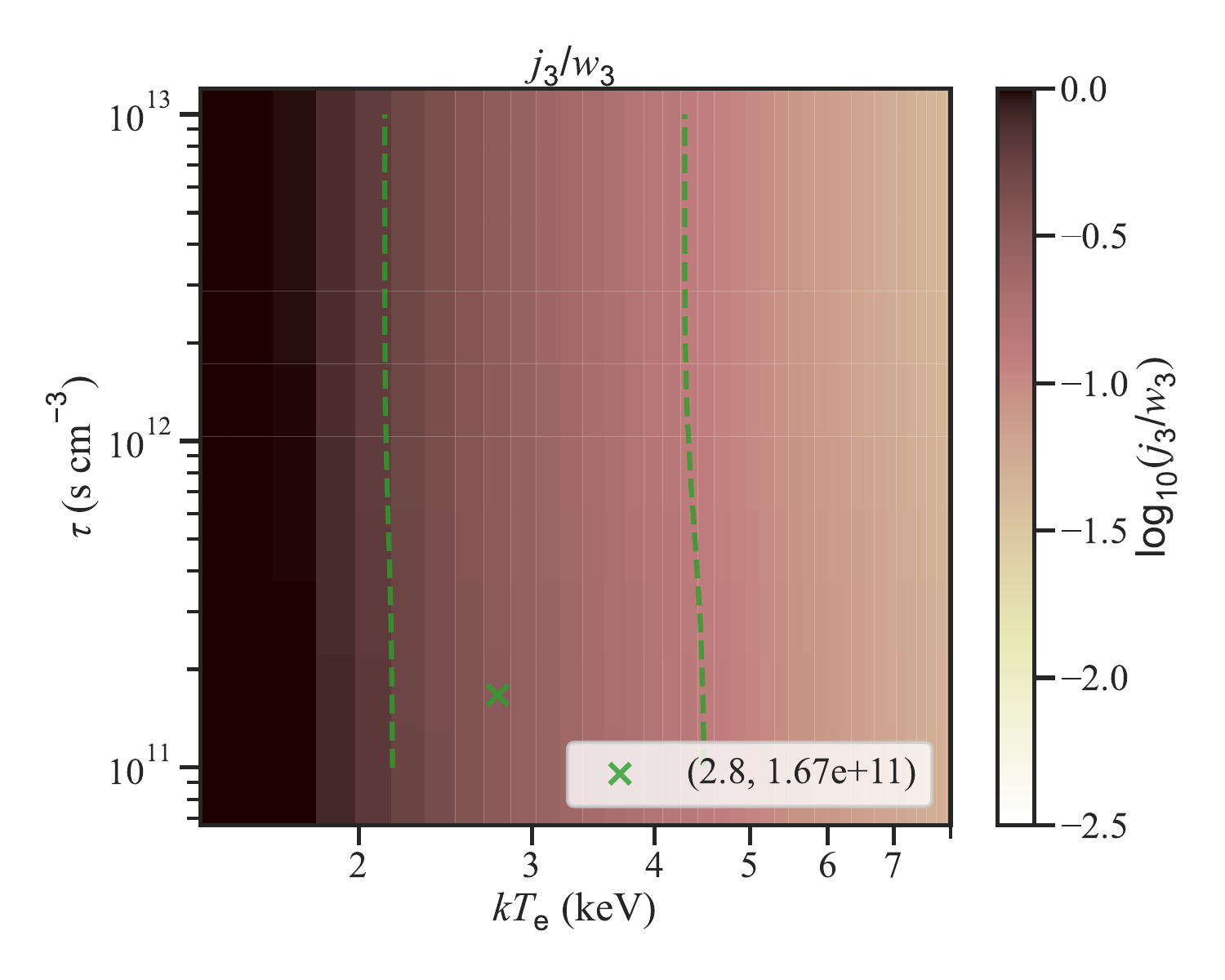}
 \includegraphics[width=0.49\linewidth]{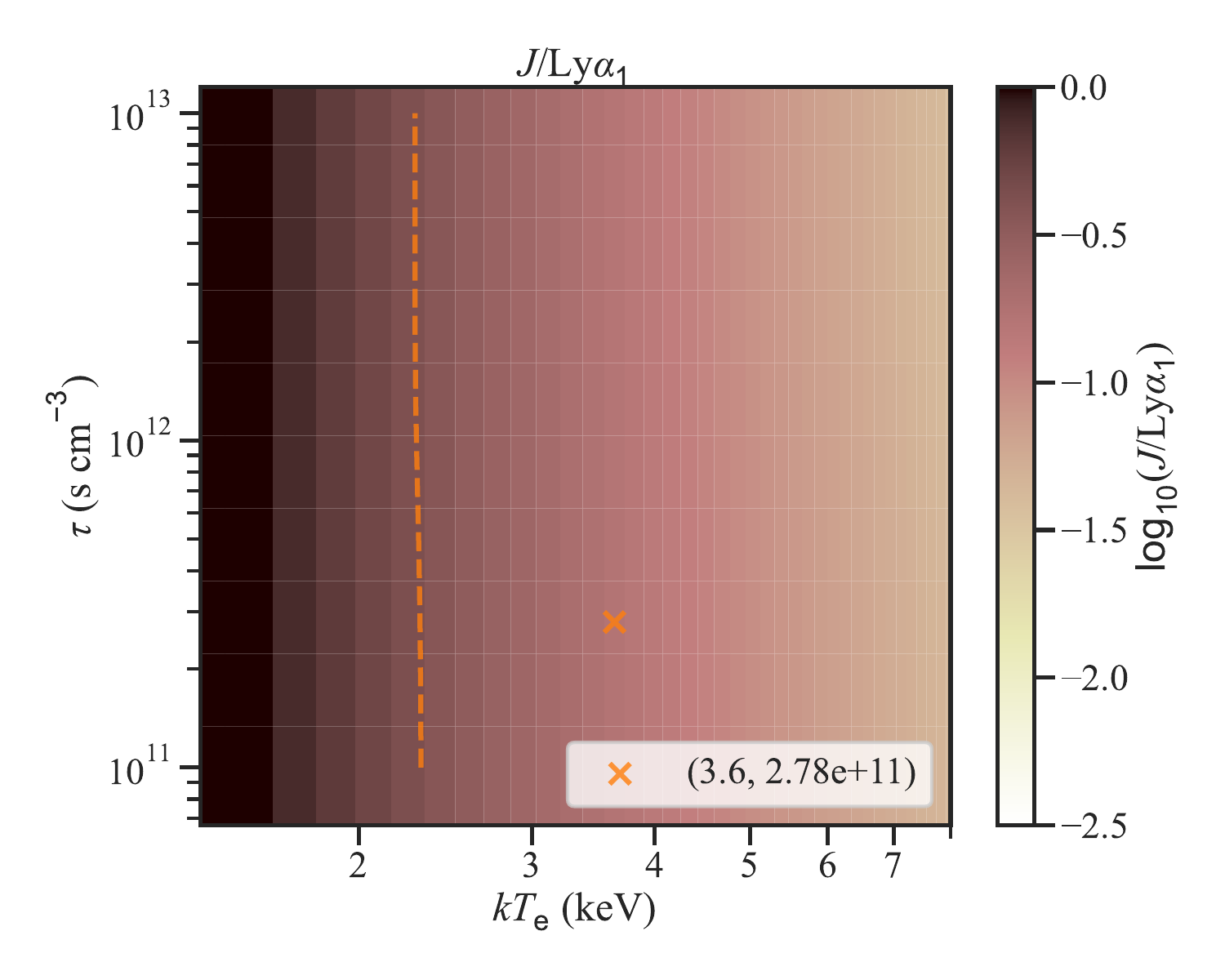}
 \includegraphics[width=0.49\linewidth]{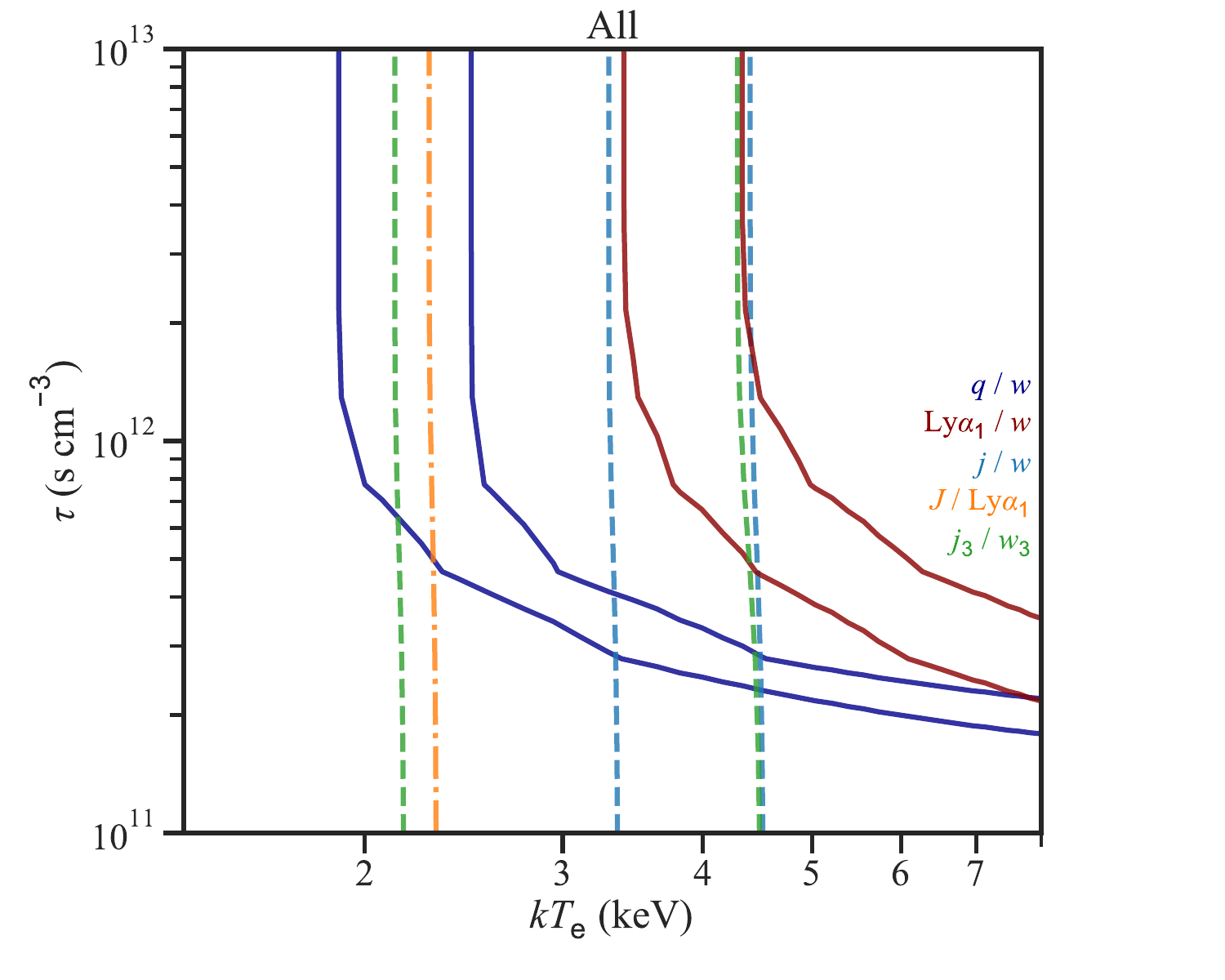}
 \end{center}
 \caption{Diagnostics for the ionizing plasma. The format and symbols follow
 \ref{f08_kappa}. The ionizing plasma model in AtomDB is retrieved via PyAtomDB
 \citep{foster2020} to calculate the five line ratios as functions of the relaxation
 parameter $\tau$ and the electron temperature $kT_{\mathrm{e}}$. The initial condition
 at $\tau=0$ assumes that all Fe ions are neutral and the electrons have a temperature
 of $kT_{\mathrm{e}}$. No space satisfies all the contours simultaneously.
 Alt text: a six-panel figure of heatmaps showing $kT_{\mathrm{e}}$ on the x-axis and
 $\tau$ on the y-axis.
 }
 \label{f08_tau}
\end{figure*}

\begin{figure*}[!hbtp]
 \begin{center}
 \includegraphics[width=0.49\linewidth]{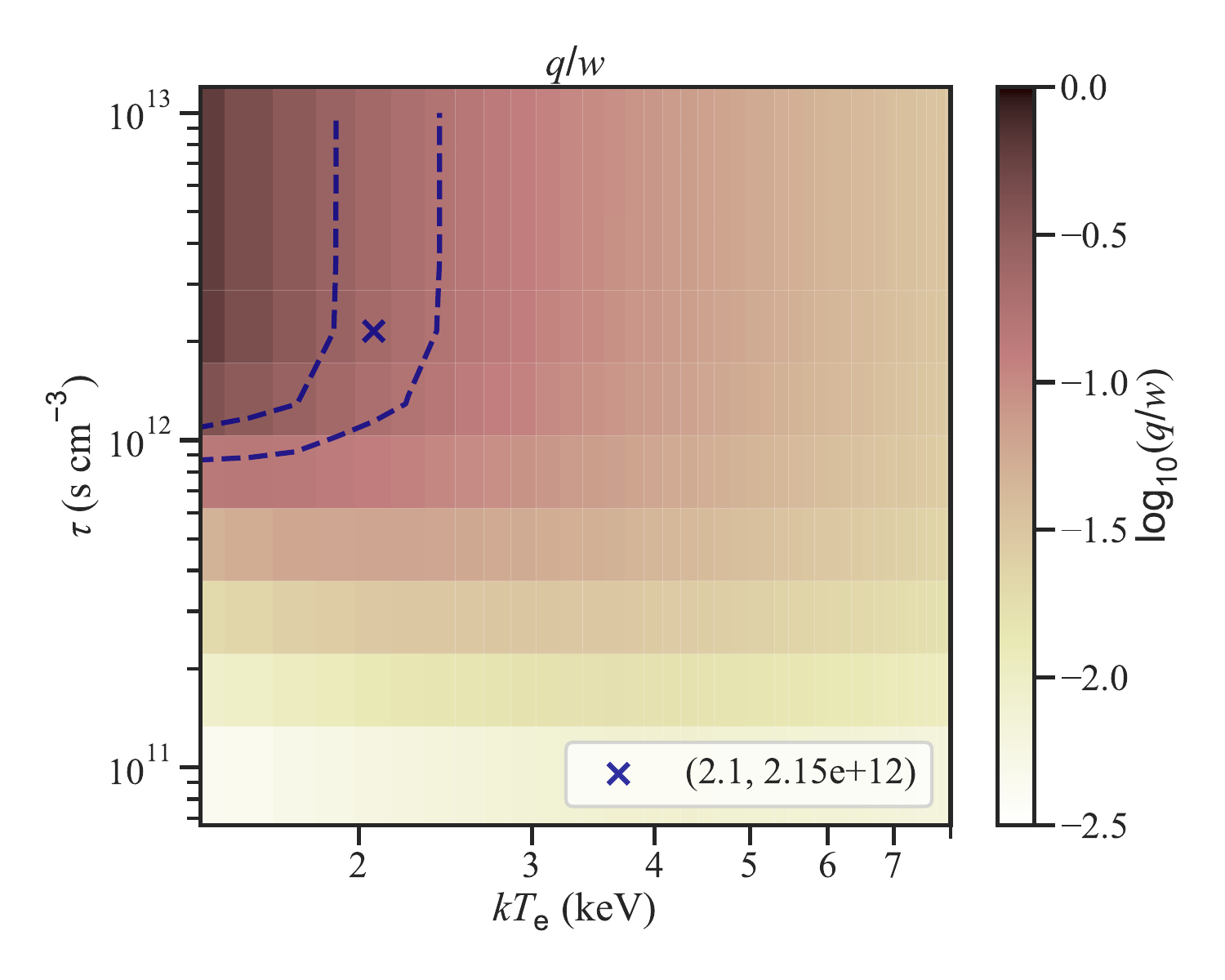}
 \includegraphics[width=0.49\linewidth]{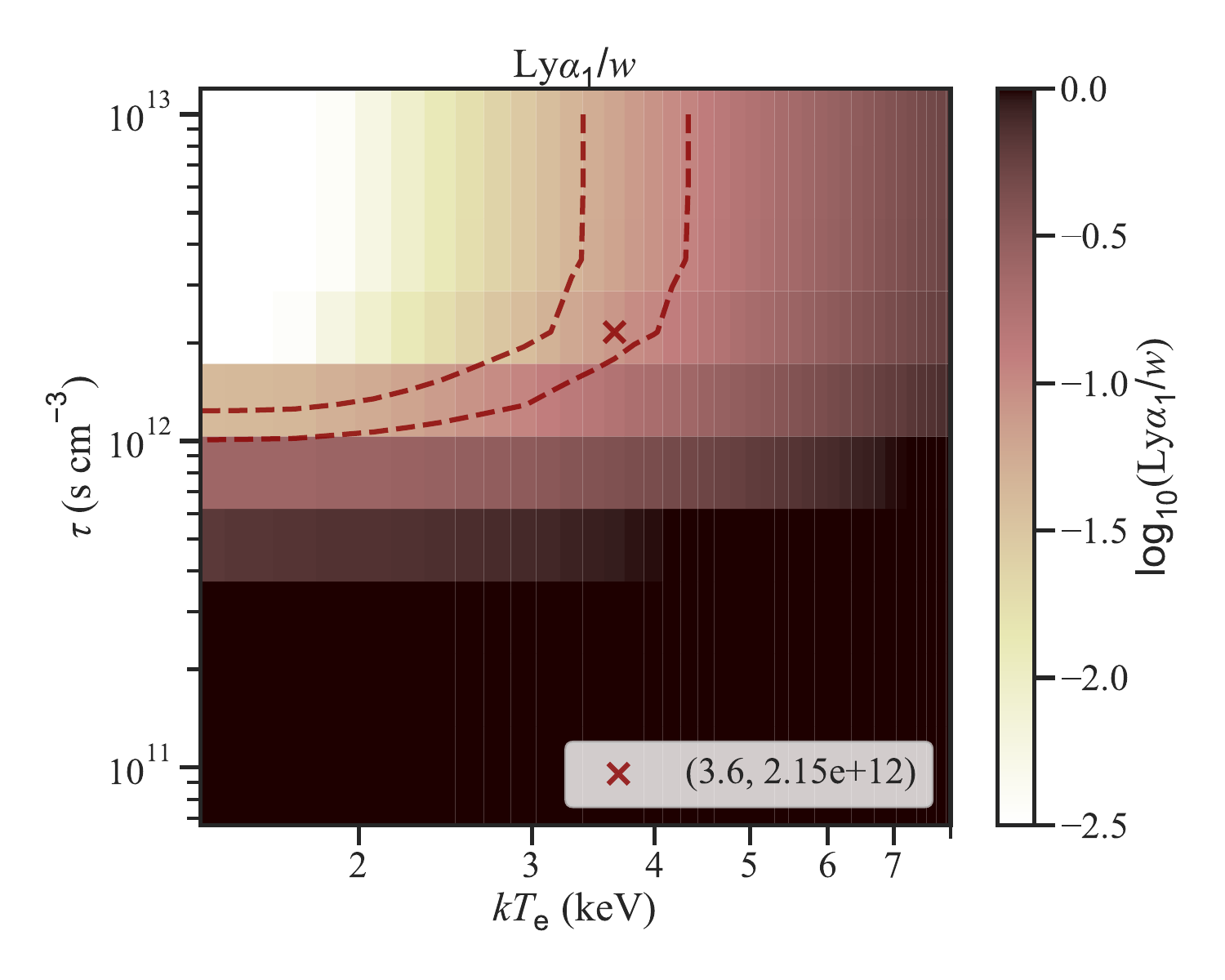}
 \includegraphics[width=0.49\linewidth]{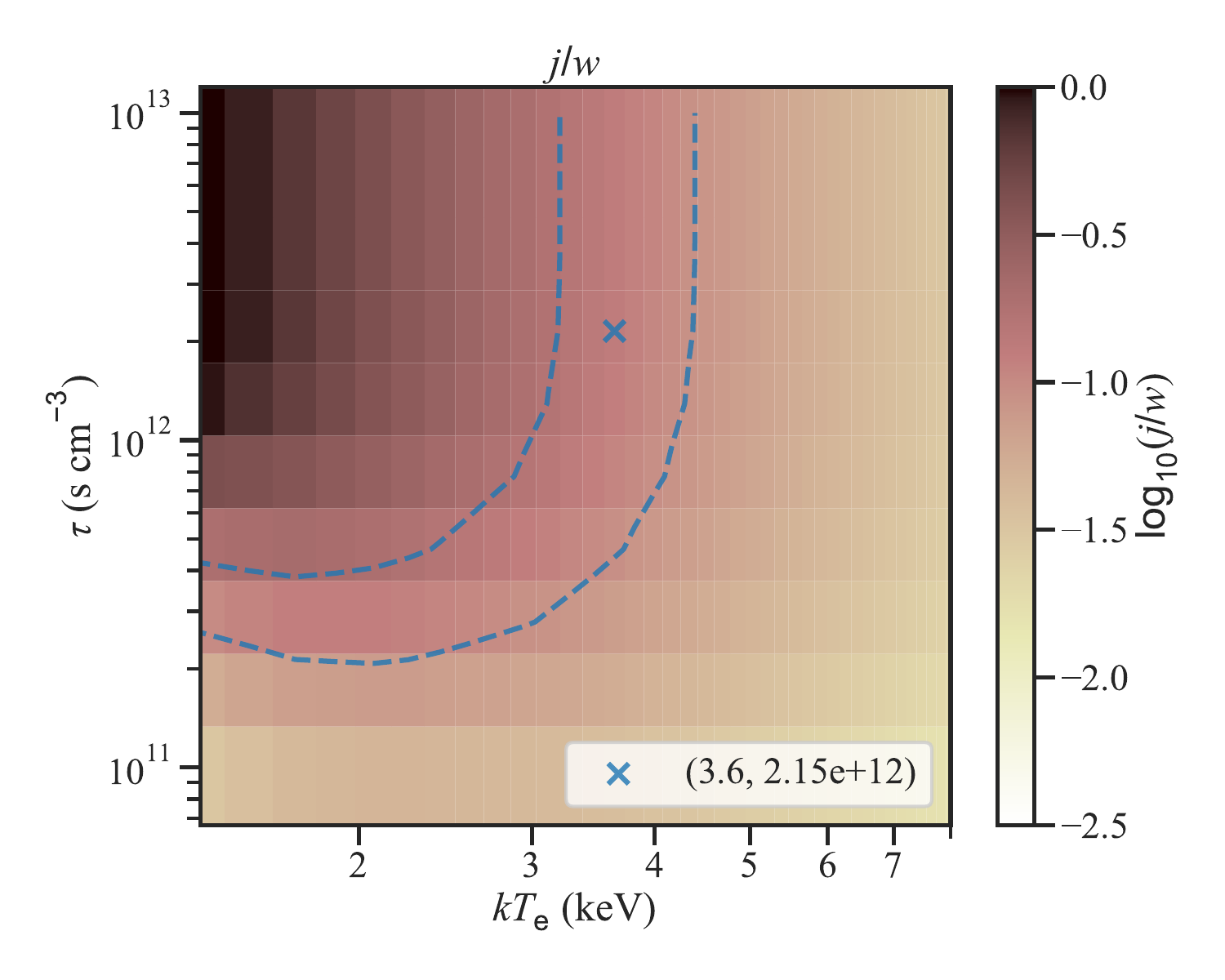}
 \includegraphics[width=0.49\linewidth]{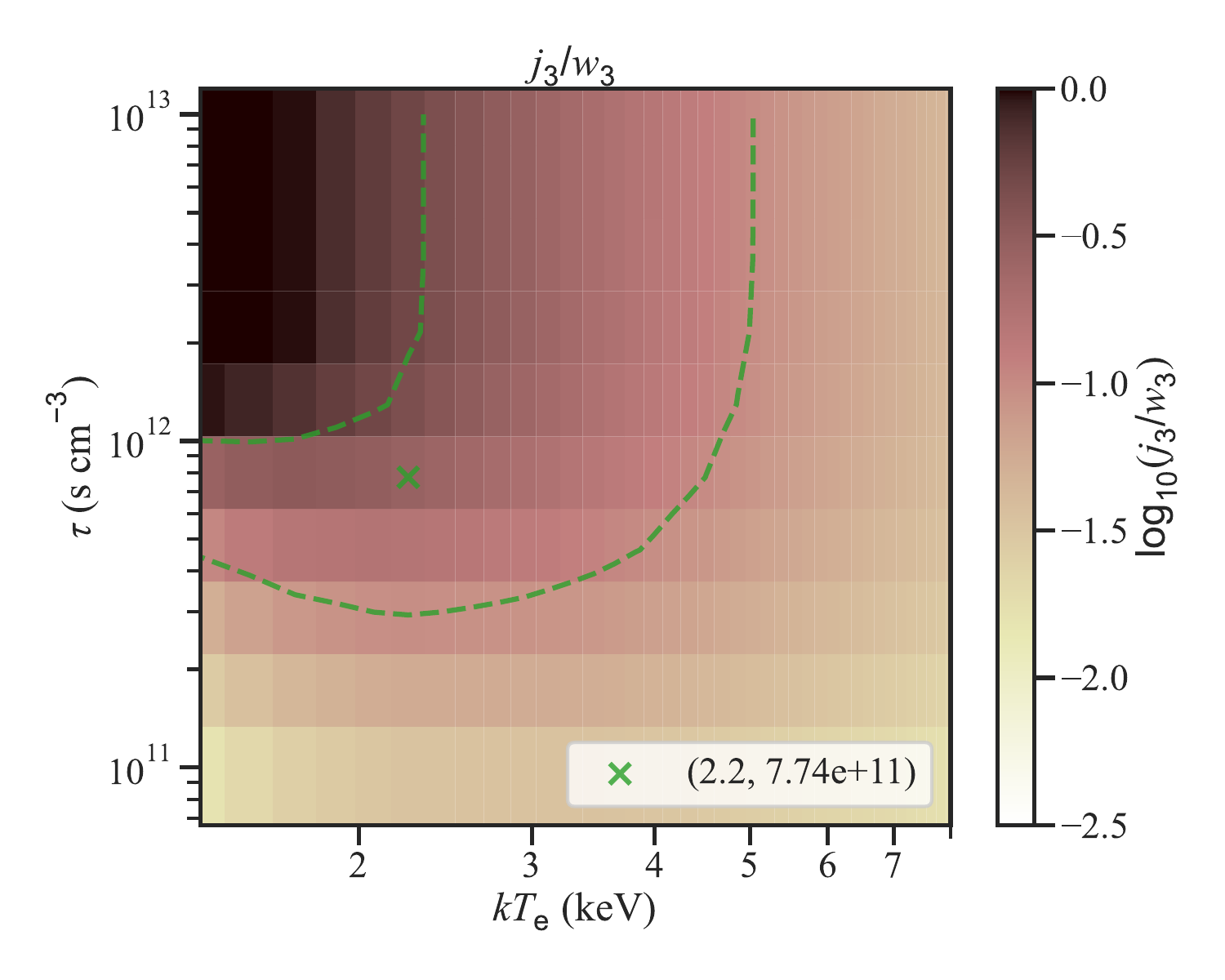}
 \includegraphics[width=0.49\linewidth]{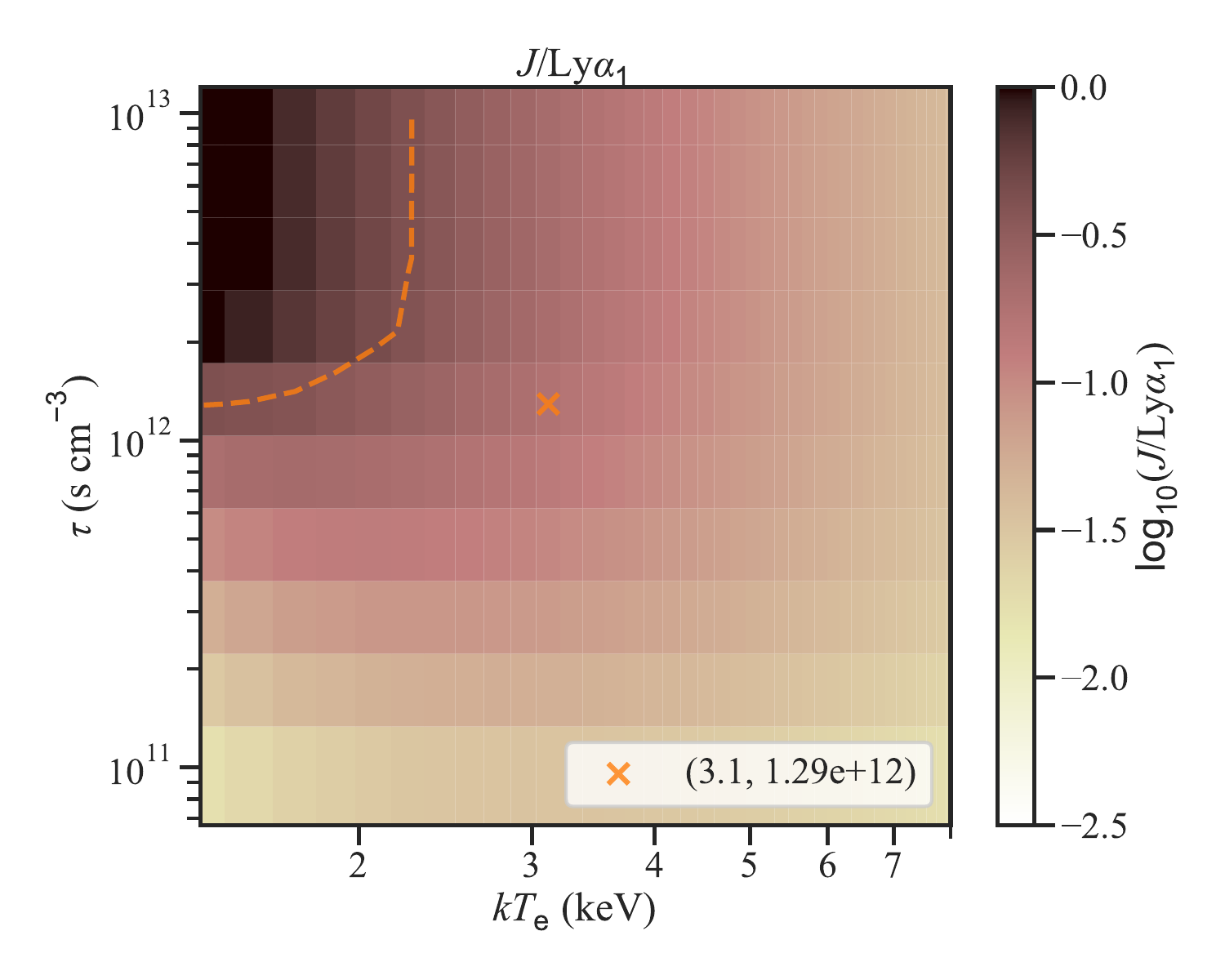}
 \includegraphics[width=0.49\linewidth]{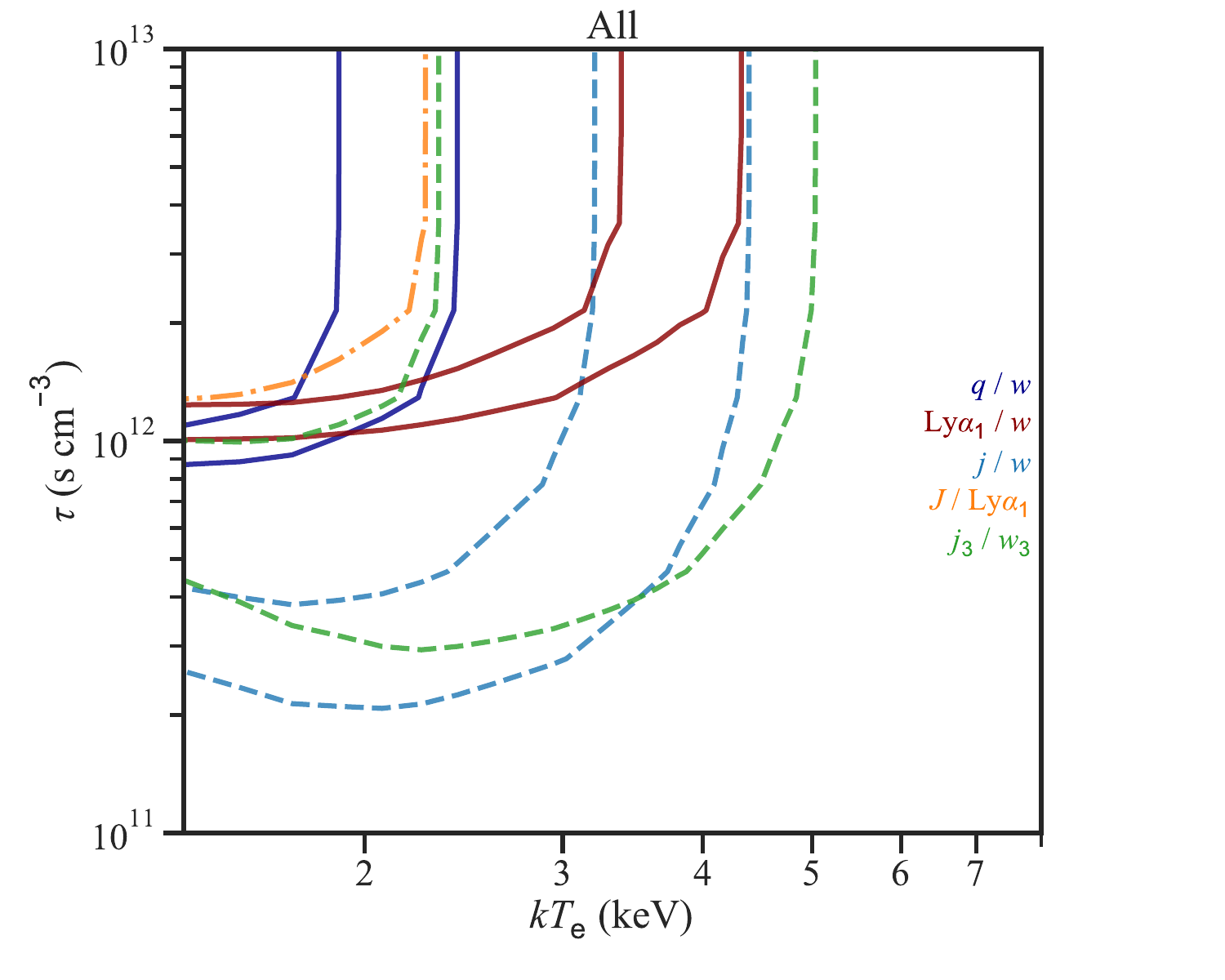}
 \end{center}
 \caption{Diagnostics for the recombining plasma. The format and symbols follow
 \ref{f08_kappa}. The ionizing plasma model in AtomDB is retrieved via PyAtomDB
 \citep{foster2020} to calculate the five
 line ratios as functions of the relaxation parameter $\tau$ and the electron temperature
 $kT_{\mathrm{e}}$. The initial condition at $\tau=0$ assumes that all Fe ions are fully
 ionized and the electrons have a temperature of $kT_{\mathrm{e}}$. No space satisfies all the contours simultaneously.
 Alt text: a six-panel figure of heatmaps showing $kT_{\mathrm{e}}$ on the x-axis and $\tau$ on the y-axis.
 }
 \label{f08_tau_rp}
\end{figure*}

The diagnostics for the ionizing and recombining plasmas are shown in
figures~\ref{f08_tau} and \ref{f08_tau_rp}, respectively. We access the NEI model based
on \texttt{AtomDB} via the \texttt{PyAtomDB} interface \citep{foster2020} to calculate
the line ratios. For the ionizing plasma, all ions start with neutral states at
$\tau=0$. For the recombining plasma, all ions start in the fully ionized state at
$\tau=0$. Both $kT_{\mathrm{e}}$ and $n_{\mathrm{e}}$ are assumed to remain constant. In
both cases, no parameter space was found that satisfies all five line ratio constraints,
suggesting that these models are not likely solutions. It is reasonable that the plasma
reaches some equilibrium, particularly for the stellar coronae at quiescence like the
one presented here.


\section{Conclusion}\label{s5}
With the unprecedented spectral resolution of XRISM $\textit{Resolve}$, detailed spectra
for Fe K-shell lines were obtained for the stellar corona from GT Mus at its quiescent
state. The satellite lines of Fe \emissiontype{XXIV}--\emissiontype{XXVI}, including
those in the Fe He$\beta$ complex, have become accessible for the first time for stellar
sources other than the Sun. 

We performed a phenomenological model fitting for the lines in three line complexes. Two
pairs of DE and three pairs of DR / DE lines were used for the line ratio analysis:
$q$/$w$, Ly$\alpha_1$/$w$, $j$/$w$, $J$/Ly$\alpha1$, and $j_3$/$w_3$. Lines in each
pair have close central energies, providing robust diagnostic results against
uncertainties in the effective area across the broad energy band.  The temperature
diagnostics through the ion charge population and the electron energy distribution
revealed that a single-temperature plasma description is insufficient, requiring
deviation from it. We investigated the two-temperature, the $\kappa$ distributions, the
ionizing or recombining plasmas, and found that the two-temperature (1.7 and 4.3~keV)
solution is preferred, which is also consistent with the broadband fitting in the
1.7--10 keV and the thermal broadening of the Fe \emissiontype{XXV} He$\alpha$ complex.

This study demonstrates the capability of the X-ray microcalorimeter spectra to diagnose
plasmas in celestial stellar sources. The same technique is applicable to spectra during
flares, which may reveal interesting non-equilibrium pheonomena. Even for the Sun, there
are only limited results in EUV (\cite{dzifvcakova2018,kawate2016}) and X-rays
\citep{doschek1987} for such features. Many celestial stellar sources are known to
exhibit gigantic and long-lasting flares compared to the Sun \citep{tsuboi2016}, and
their observations with \textit{Resolve} are awaited.

\begin{ack}
We appreciate insightful discussions on high-resolution X-ray spectroscopy with Tetsuya
Watanabe at NAOJ.  This research has made use of data and/or software provided by the
High Energy Astrophysics Science Archive Research Center (HEASARC), which is a service
of the Astrophysics Science Division at NASA/GSFC. \texttt{Chianti} is a collaborative
project involving George Mason University, the University of Michigan (USA), University
of Cambridge (UK) and NASA Goddard Space Flight Center (USA). This research was
supported by the grant of Joint Research by the National Institutes of Natural Sciences
(NINS) (NINS program No OML032402). The material is based upon work supported by NASA
under award number 80GSFC21M0002. EB acknowledges support from NASA grants
80NSSC20K0733, 80NSSC24K1148, and 80NSSC24K1774. Part of this work was performed under
the auspices of the U.S. Department of Energy by Lawrence Livermore National Laboratory
under Contract DE-AC52-07NA27344.
\end{ack}

\bibliographystyle{aa}
\bibliography{ms1}
\end{document}